\documentclass[12pt]{article}
\usepackage{axodraw,amsmath,amssymb,amsfonts,color,graphicx,cite,color,feynarts}
\input paperdef

\graphicspath{{figsmini/}}

\evensidemargin \oddsidemargin
\allowdisplaybreaks

\begin{document}
\thispagestyle{empty}

\def\thefootnote{\fnsymbol{footnote}}

\begin{flushright}
DCPT/06/160 \\
IPPP/06/80 \hfill
MPP--2006--158\\
PSI--PR--06--14 \hfill
hep-ph/0611326
\end{flushright}

\vspace{1cm}

\begin{center}

{\large\sc {\bf The Higgs Boson Masses and Mixings of the}}

\vspace{0.4cm}

{\large\sc {\bf Complex MSSM in the Feynman-Diagrammatic Approach}}

\vspace{1cm}

{\sc
M.~Frank$^{1}$%
\footnote{email: m@rkusfrank.de}%
, T.~Hahn$^{2}$%
\footnote{email: hahn@feynarts.de}%
, S.~Heinemeyer$^{3}$%
\footnote{email: Sven.Heinemeyer@cern.ch}%
, W.~Hollik$^{2}$%
\footnote{email: hollik@mppmu.mpg.de}%
,\\[.2em] H.~Rzehak$^{4}$%
\footnote{email: Heidi.Rzehak@psi.ch}%
~and G.~Weiglein$^{5}$%
\footnote{email: Georg.Weiglein@durham.ac.uk}
}

\vspace*{.7cm}

{\sl
$^1$Institut f\"ur Theoretische Physik, Universit\"at Karlsruhe, \\
D--76128 Karlsruhe, Germany\footnote{former address}

\vspace*{0.1cm}

$^2$Max-Planck-Institut f\"ur Physik (Werner-Heisenberg-Institut),
F\"ohringer Ring 6, \\
D--80805 M\"unchen, Germany

\vspace*{0.1cm}

$^3$Instituto de Fisica de Cantabria (CSIC-UC), Santander,  Spain

\vspace*{0.1cm}

$^4$Paul Scherrer Institut, W\"urenlingen und Villigen, CH--5232
Villigen PSI, Switzerland

\vspace*{0.1cm}

$^5$IPPP, University of Durham, Durham DH1~3LE, UK
}

\end{center}

\vspace*{0.1cm}

\begin{abstract}
\noindent
New results for the complete one-loop contributions to the masses and 
mixing effects in the Higgs sector are obtained for the MSSM with
complex parameters using the Feynman-diagrammatic approach. 
The full dependence on all relevant complex phases
is taken into account, and all the imaginary parts appearing in the
calculation are treated in a consistent way. 
The renormalization is discussed in detail, and a hybrid 
on-shell/\DRbar\ scheme is adopted.
We also derive the wave function normalization factors needed in 
processes with external Higgs bosons and discuss effective couplings 
incorporating leading higher-order effects.
The complete \onel\ corrections, supplemented by the available
two-loop corrections in the Feynman-diagrammatic approach for the 
MSSM with real parameters and a resummation of the leading (s)bottom
corrections for complex parameters, are implemented into the public Fortran
code \fhtt.
In our numerical analysis the full results
for the Higgs-boson masses and couplings are compared with various
approximations, and $\cp$-violating effects in the mixing of the heavy Higgs
bosons are analyzed in detail. We find sizable deviations in comparison
with the approximations often made in the literature.
\end{abstract}

\def\thefootnote{\arabic{footnote}}
\setcounter{page}{0}
\setcounter{footnote}{0}

\newpage


\section{Introduction}

A striking prediction of models of 
supersymmetry (SUSY)~\cite{mssm} is a Higgs sector with at
least one relatively light Higgs boson. In the Minimal Supersymmetric
extension of the Standard Model (MSSM) two Higgs doublets are required,
resulting in five physical Higgs bosons: the light and heavy $\cp$-even $h$
and $H$, the $\cp$-odd $A$, and the charged Higgs bosons $H^\pm$.
The Higgs sector of the MSSM can be expressed at lowest
order in terms of $\MZ$ (or $\MW$), $\MA$ (or $\MHp$) and 
$\tb \equiv v_2/v_1$, 
the ratio of the two vacuum expectation values. All other masses and
mixing angles can therefore be predicted. Higher-order contributions give 
large corrections to the tree-level relations.

The limits obtained from the Higgs search at LEP (the final LEP
results can be found in \citeres{LEPHiggsSM,LEPHiggsMSSM}), place
important restrictions on the parameter space of the MSSM.
The results obtained so far 
at Run~II of the Tevatron~\cite{D0bounds,CDFbounds,Tevcharged}
yield interesting constraints in particular in the region of small $\MA$
and large $\tan\beta$ (the dependence on the other MSSM parameters has
recently been analyzed in \citere{benchmark3}). The Large Hadron Collider 
(LHC) has good prospects for the discovery of
at least one Higgs boson over all the 
MSSM parameter space~\cite{lhctdrs,atlashiggs,cmshiggs} 
(see e.g.\ \citeres{jakobs,schumi} 
for recent reviews). At the International Linear Collider (ILC)
eventually high-precision physics in the Higgs sector may become
possible~\cite{tesla,orangebook,acfarep}.
The interplay of the LHC and the ILC in the MSSM Higgs sector is
discussed in \citeres{lhcilc,eili}.

For the MSSM with real parameters (rMSSM) the status of higher-order
corrections to the masses and mixing angles in the Higgs sector 
is quite advanced. The complete one-loop result within the rMSSM is
known~\cite{ERZ,mhiggsf1lA,mhiggsf1lB,mhiggsf1lC}.
The by far dominant
one-loop contribution is the \order{\alt} term due to top and stop
loops ($\alt \equiv h_t^2 / (4 \pi)$, $h_t$ being the
top-quark Yukawa coupling). The computation of the two-loop corrections
has meanwhile reached a stage where all the presumably dominant
contributions are 
available~\cite{mhiggsletter,mhiggslong,mhiggslle,mhiggsFD2,bse,mhiggsEP0,mhiggsEP1,mhiggsEP1b,mhiggsEP2,mhiggsEP3,mhiggsEP3b,mhiggsEP4,mhiggsEP4b,mhiggsRG1,mhiggsRG1a}.
In particular, the \order{\alt\als}, \order{\alt^2}, \order{\alb\als},
\order{\alt\alb} and \order{\alb^2} contributions to the self-energies
are known for vanishing external momenta.  For the (s)bottom
corrections, which are mainly relevant for large values of $\tb$,
an all-order resummation of the $\tb$-enhanced term of
\order{\alb(\als\tb)^n} is
performed~\cite{deltamb1,deltamb2,deltamb2b,deltamb3}. 
The remaining theoretical uncertainty on the lightest $\cp$-even Higgs
boson mass has been estimated to be below 
$\sim 3 \gev$~\cite{mhiggsAEC,PomssmRep,mhiggsWN}. 
The above calculations have been implemented into public 
codes. The program \fh~\cite{mhiggslong,feynhiggs,feynhiggs1.2,feynhiggs2}
is based on the results obtained in the Feynman-diagrammatic (FD)
approach~\cite{mhiggsletter,mhiggslong,mhiggsAEC,mhiggsEP4b}. It
includes all the above corrections. 
The code {\tt CPsuperH}~\cite{cpsh} is based on the renormalization group (RG)
improved effective potential approach~\cite{mhiggsRG1a,mhiggsRG1,bse}.
Most recently a full two-loop effective potential calculation%
\footnote{
In \citere{dredDS2} the symmetry relations affecting higher-order
corrections in the MSSM Higgs sector have been analyzed in detail. It 
has been shown for
those two-loop corrections that are implemented in
\fhtt\ that the counterterms arising from multiplicative
renormalization preserve SUSY, so that the existing result is valid
without the introduction of additional symmetry-restoring counterterms. 
It is not yet clear whether the same is true also for the subleading
two-loop corrections obtained in \citere{mhiggsEP5}.
}
(including even the momentum dependence for the leading
pieces) has been published~\cite{mhiggsEP5}. However, no computer code
is publicly available.
Besides the masses in the Higgs sector, 
also for the couplings of the rMSSM Higgs bosons to SM bosons and
fermions detailed higher-order corrections are
available~\cite{hff,deltamb1,deltamb2,deltamb2b,higgscoup}.

In the case of the MSSM with complex parameters (cMSSM) the higher
order corrections have yet been restricted, after the first more
general investigations~\cite{mhiggsCPXgen}, to evaluations in the
effective potential (EP)
approach~\cite{mhiggsCPXEP,mhiggsCPXsn} (at one-loop, neglecting the
momentum-dependent effects) and to the RG improved \onel\ EP 
method~\cite{mhiggsCPXRG1,mhiggsCPXRG2}. The latter ones have been
restricted to the corrections arising from the (s)fermion sector and
some leading logarithmic corrections from the gaugino sector%
\footnote{
The two-loop results of \citere{mhiggsEP5} can in principle also be
taken over to the cMSSM. However, no explicit evaluation or computer
code based on these results exists.
}%
. Within the FD approach the \onel\ leading $\mt^4$
corrections 
have been evaluated in \citere{mhiggsCPXFD1}. 
Effects of imaginary parts of the one-loop contributions to
Higgs boson masses and couplings 
have been considered in \citeres{imagSE1,imagSE2,imagSE3}.
Further discussions on the effect of complex phases on Higgs boson
masses and decays can be found in 
\citeres{mhcMSSM2L,cpnshHiggs,ll06sh,susy06hr}.
A detailed
comparison between the two available computer codes for the cMSSM Higgs-boson
sector, \fh\ and {\tt CPsuperH}, will be performed in a
forthcoming publication.

In the present paper we present the complete one-loop evaluation
of the Higgs-boson masses and mixings in the cMSSM (see 
\citere{mhiggsCPXFDproc} for preliminary results).
The full phase dependence, the full momentum dependence and the
effects of imaginary parts of the Higgs-boson self-energies are taken
consistently into account. Our results 
are based on the FD approach using a hybrid renormalization scheme where
the masses are renormalized on-shell, while the \DRbar\ scheme is
applied for $\tb$ and the field renormalizations. 
The higher-order self-energy corrections are utilized to obtain wave
function normalization factors for external Higgs bosons and to discuss
effective couplings incorporating
leading higher-order effects.
We provide numerical examples for the lightest
cMSSM Higgs boson, the mass difference of the heavier neutral Higgses
and for the mixing between the three neutral Higgs bosons. 
We compare our results with various approximations often made in the
literature.
All results are incorporated into the public Fortran code
\fhtt~\cite{mhiggslong,feynhiggs,feynhiggs1.2,feynhiggs2}. 

The rest of the paper is organized as follows. 
In \refse{sec:calc} we review all relevant sectors of the cMSSM.
Besides the tree-level structure of the Higgs sector, the
renormalization necessary for the one-loop calculations is explained
in detail. In \refse{sec:higherorderhiggs} the evaluation of the
one-loop self-energies is presented. 
The determination of the Higgs-boson masses from the propagators and of 
wave function normalization factors and effective couplings is
described. 
Our numerical analysis is given in
\refse{sec:numeval}. Information about the Fortran code \fhtt\ is
provided in \refse{sec:feynhiggs}, more details about installation and
use are given in the Appendix. We conclude with
\refse{sec:conclusions}. 


\section{Calculational basis}
\label{sec:calc}

\subsection{The scalar quark sector in the cMSSM}

The mass matrix of two squarks of the same flavor, $\sql$ and $\sqr$,
is given by 
\begin{align}
M_{\sq} =
\begin{pmatrix}
        M_L^2 + \mq^2 + \MZ^2 \CZb (I_3^q - Q_q \sw^2) & \mq \; \Xq^* \\
        \mq \; \Xq    & M_{\tilde{q}_R}^2 + \mq^2 + \MZ^2 \CZb Q_q \sw^2
\end{pmatrix} ,
\label{squarkmassmatrix}
\end{align}
with
\BEA
\Xq &=& A_q - \mu^* \{\CTb, \tb\} ,
\label{squarksoftSUSYbreaking}
\EEA
where $\{\CTb, \tb\}$ applies for up- and down-type squarks, respectively,
the star denotes a complex conjugation,
and $\tb \equiv v_2/v_1$.
In \refeq{squarksoftSUSYbreaking}
$M_L^2$, $M_{\tilde{q}_R}^2$
are real soft SUSY-breaking
parameters, while the soft SUSY-breaking trilinear coupling $A_q$ and
the higgsino mass parameter $\mu$ can be complex. As a consequence,
in the scalar quark sector of the cMSSM $N_q + 1$ phases are
present, one for each $A_q$ and one
for $\mu$, i.e.\ $N_q + 1$ new parameters appear. As an abbreviation we
will use
\BE
\varphi_{\Xq} \equiv {\rm arg}\KL \Xq \KR , \; 
\varphi_{\Aq} \equiv \arg\KL\Aq\KR~.
\end{equation}
One can trade $\varphi_{\Aq}$ for
$\varphi_{\Xq}$ as independent parameter. \\
The squark mass eigenstates are obtained by the unitary transformation
\BE
\VL \sqe \\ \sqz \VR = \matr{U}_{\sq} \VL \sql \\ \sqr \VR
\end{equation}
with
\BE
\matr{U}_{\sq} = \ML \ctq & \stq \\ -\stq^* & \ctq \MR
               ,\quad
\matr{U}_{\sq} \matr{U}_{\sq}^{\dagger} = \unity~,
\end{equation}
The elements of the mixing matrix $\matr{U}$ can be calculated as
\begin{align}
  c_{\tilde{q}} &=
  \frac{\sqrt{M_L^2 + \mq^2 + \MZ^2 \CZb (I_3^q - Q_q \sw^2) - m_{\tilde{q}_2}^2}}
       {\sqrt{m_{\tilde{q}_1}^2 - m_{\tilde{q}_2}^2}},\\[0.2cm] 
  s_{\tilde{q}} &=
  \frac{m_q \Xq^*}
       {\sqrt{M_L^2 + \MZ^2 \CZb (I_3^q - Q_q \sw^2) + \mq^2 - m_{\tilde{q}_2}^2} 
        \sqrt{m_{\tilde{q}_1}^2 - m_{\tilde{q}_2}^2}}.
\end{align}
Here $\ctq \equiv \costq$ is real, whereas 
$\stq \equiv e^{-i\varphi_{\Xq}}\,\sintq$ 
can be complex with the phase
\BE
\varphi_{\stq} = - \varphi_{\Xq} = {\rm arg}\KL \Xq^* \KR ~.
\end{equation}

The mass eigenvalues are given by
\begin{align}
m_{\tilde q_{1,2}}^2 = \mq^2
  + \edz 
&\Bigl[
M_L^2 + M_{\tilde{q}_R}^2 + I_3^q \MZ^2 \CZb \\&
           \mp \sqrt{[M_L^2 - M_{\tilde{q}_R}^2 + \MZ^2 \CZb(I_3^q -2 Q_q
  \sw^2)]^2 + 4 \mq^2 |\Xq|^2}~\Bigr],
\end{align}
and are independent of the phase of $\Xq$.


\subsection{The chargino / neutralino sector of the cMSSM}

The mass eigenstates of the charginos can be determined from the matrix
\begin{align}
  \matr{X} =
  \begin{pmatrix}
    \MTwo & \sqrt{2} \sinb \MW \\
    \sqrt{2} \cosb \MW & \mu
  \end{pmatrix}.
\end{align}
In addition to the higgsino mass parameter $\mu$ it 
contains the soft breaking term $\MTwo$, 
which can also be complex in the cMSSM.
The rotation to the chargino mass eigenstates is done by transforming
the original wino and higgsino fields with the help of two unitary 2$\times$2
matrices $\matr{U}$ and $\matr{V}$,
\begin{align}
\label{eq:charginotransform}
  \begin{pmatrix} \tilde{\chi}_1^+ \\ \tilde{\chi}_2^+ \end{pmatrix} =
  \matr{V}
  \begin{pmatrix} \tilde{W}^+ \\ \tilde{H}_2^+ \end{pmatrix}, \quad
  \begin{pmatrix} \tilde{\chi}_1^- \\ \tilde{\chi}_2^- \end{pmatrix} =
  \matr{U}
  \begin{pmatrix} \tilde{W}^- \\ \tilde{H}_1^- \end{pmatrix} .
\end{align}
These rotations lead to the diagonal mass matrix
\begin{align}
  \begin{pmatrix} m_{\tilde{\chi}^\pm_1} & 0 \\ 0 &
    m_{\tilde{\chi}^\pm_2} \end{pmatrix} = 
  \matr{U}^* \, \matr{X} \, \matr{V}^{\dagger}.
\end{align}
{}From this relation, it becomes clear that the chargino masses
$m_{\tilde{\chi}^\pm_1}$ and
$m_{\tilde{\chi}^\pm_2}$ can be determined as the (real and positive)
singular values of $\matr{X}$.
The singular value decomposition of $\matr{X}$
also yields results for
$\matr{U}$ and~$\matr{V}$.

A similar procedure is used for the determination of the neutralino masses and
mixing matrix, which can both be calculated from the mass matrix
\begin{align}
  \matr{Y} =
  \begin{pmatrix}
    \MOne                  & 0                & -\MZ \, \sw \cosb
    & \MZ \, \sw \sinb \\ 
    0                      & \MTwo            & \quad \MZ \, \cw \cosb
    & \MZ \, \cw \sinb \\ 
    -\MZ \, \sw \cosb      & \MZ \, \cw \cosb & 0
    & -\mu             \\ 
    \quad \MZ \, \sw \sinb & \MZ \, \cw \sinb & -\mu                    & 0
  \end{pmatrix}.
\end{align}
This symmetric matrix contains the additional complex soft-breaking
parameter $\MOne$. 
The diagonalization of the matrix
is achieved by a transformation starting from the original
bino/wino/higgsino basis,
\begin{align}
  \begin{pmatrix} \tilde{\chi}^0_1 \\[.5em] \tilde{\chi}^0_2 \\[.5em]
    \tilde{\chi}^0_3 \\[.5em] \tilde{\chi}^0_4 \end{pmatrix} = 
  \matr{N}
  \begin{pmatrix} \tilde{B}^0 \\[.5em] \tilde{W}^0 \\[.5em] 
  \tilde{H}_1^0 \\[.5em]  \tilde{H}_2^0 \end{pmatrix}, \quad 
  \begin{pmatrix}
    m_{\tilde{\chi}^0_1} & 0 & 0 & 0 \\[.5em]
    0 & m_{\tilde{\chi}^0_2} & 0 & 0 \\[.5em]
    0 & 0 & m_{\tilde{\chi}^0_3} & 0 \\[.5em]
    0 & 0 & 0 & m_{\tilde{\chi}^0_4}
  \end{pmatrix} =
  \matr{N}^* \, \matr{Y} \, \matr{N}^{\dagger}.
\end{align}
The unitary 4$\times$4 matrix $\matr{N}$ and the physical neutralino
masses again result 
from a numerical singular value decomposition of $\matr{Y}$.
The symmetry of $\matr{Y}$ permits the non-trivial condition of
using only one 
matrix $\matr{N}$ for its diagonalization, in contrast to the chargino
case shown above.


\subsection{The cMSSM Higgs potential}

The Higgs potential $\VHiggs$ contains the real soft breaking terms
$\tilde m_1^2$ and $\tilde m_2^2$ (with $m_1^2 \equiv \tilde m_1^2 +|\mu|^2$,
$m_2^2 \equiv \tilde m_2^2 + |\mu|^2$), 
the potentially complex soft breaking parameter
$m_{12}^2$, and the U$(1)$ and SU$(2)$ coupling constants $g_1$ and $g_2$: 
\begin{align}
\label{eq:higgspotential}
  \VHiggs &= m_1^2 H_{1i}^* H_{1i} + m_2^2 H_{2i}^* H_{2i} 
- \epsilon^{ij} 
(m_{12}^2 H_{1i} H_{2j} + {m_{12}^2}^* H_{1i}^*
  H_{2j}^*) \notag \\[.5em]
  &\quad +\tfrac{1}{8}(g_1^2+g_2^2)(H_{1i}^* H_{1i} - H_{2i}^* H_{2i})^2 
    + \tfrac{1}{2} g_2^2 | H_{1i}^* H_{2i} |^2. 
\end{align}

\noindent
The indices $\{i,j\}=\{1,2\}$ refer to the respective Higgs doublet
component (summation over $i$ and $j$ is understood), 
and $\epsilon^{12}= 1$. 
The Higgs doublets are decomposed in the following way,
\begin{align}
\label{eq:higgsdoublets}
\cHe = \begin{pmatrix} H_{11} \\ H_{12} \end{pmatrix} &=
\begin{pmatrix} v_1 + \tfrac{1}{\sqrt{2}} (\phi_1-i \chi_1) \\
  -\phi^-_1 \end{pmatrix}, \notag \\ 
\cHz = \begin{pmatrix} H_{21} \\ H_{22} \end{pmatrix} &= e^{i \xi}
\begin{pmatrix} \phi^+_2 \\ v_2 + \tfrac{1}{\sqrt{2}} (\phi_2+i
  \chi_2) \end{pmatrix}. 
\end{align}
Besides the vacuum expectation values $v_1$ and $v_2$,
eq.~(\ref{eq:higgsdoublets}) introduces a 
possible new phase $\xi$ between the two Higgs doublets. 
Using this decomposition, $\VHiggs$ can be rearranged in powers of the fields,
\begin{align}
\VHiggs &=  \cdots - T_{\phi_1} \phi_1 - T_{\phi_2} \phi_2 -
        T_{\chi_1} \chi_1 - T_{\chi_2} \chi_2 + \notag \\ 
        &+ \tfrac{1}{2} \begin{pmatrix} \phi_1,\phi_2,\chi_1,\chi_2
        \end{pmatrix} 
\matr{M}_{\phi\phi\chi\chi}
\begin{pmatrix} \phi_1 \\ \phi_2 \\ \chi_1 \\ \chi_2 \end{pmatrix} +
\begin{pmatrix} \phi^-_1,\phi^-_2  \end{pmatrix}
\matr{M}_{\phi^\pm\phi^\pm}
\begin{pmatrix} \phi^+_1 \\ \phi^+_2  \end{pmatrix} + \cdots,
\end{align}
where the coefficients of the linear terms are called tadpoles and
those of the bilinear terms are the mass matrices
$\matr{M}_{\phi\phi\chi\chi}$ and $\matr{M}_{\phi^\pm\phi^\pm}$. 
The tadpole coefficients read
\begin{subequations}
\label{eq:phichi_tadpoles}
\begin{align}
\label{eq:phi1_tadpole}
T_{\phi_1} &= -\sqrt{2} (m_1^2 v_1 - \cos \xi' |m_{12}^2| v_2 +
\tfrac{1}{4}(g_1^2 + g_2^2)(v_1^2 - v_2^2) v_1), \\ 
\label{eq:phi2_tadpole}
T_{\phi_2} &= -\sqrt{2} (m_2^2 v_2 - \cos \xi' |m_{12}^2| v_1 -
\tfrac{1}{4}(g_1^2 + g_2^2)(v_1^2 - v_2^2) v_2), \\ 
\label{eq:chi1_tadpole}
T_{\chi_1} &= \sqrt{2} \sin \xi' |m_{12}^2| v_2 = - T_{\chi_2}
\frac{v_2}{v_1},
\end{align}
\end{subequations}
with $\xi' \equiv \xi + \arg(m_{12}^2)$.

With the help of a Peccei-Quinn
transformation~\cite{Peccei} $\mu$ and $m_{12}^2$ can be 
redefined~\cite{MSSMcomplphasen} such that the complex phase of 
$m_{12}^2$ vanishes.
In the following we will therefore treat $m_{12}^2$ as a real
parameter, which yields
\begin{equation}
|m_{12}^2| = m_{12}^2, \quad \xi' = \xi .
\label{eq:pecceiquinn}
\end{equation}

\smallskip
The real, symmetric 4$\times$4-matrix 
$\matr{M}_{\phi\phi\chi\chi}$ and the hermitian 2$\times$2-matrix 
$\matr{M}_{\phi^\pm\phi^\pm}$ contain the following elements, 
\begin{subequations}
\begin{align}
\matr{M}_{\phi\phi\chi\chi} &=
\begin{pmatrix} \matr{M}_\phi                 & \matr{M}_{\phi \chi} \\[.5em] 
                \matr{M}_{\phi\chi}^{\dagger} & \matr{M}_{\chi} 
\end{pmatrix}, 
\end{align}

\begin{align}
\label{eq:MphiKomponenten}
\matr{M}_\phi  &= 
\begin{pmatrix}
m_1^2 + \tfrac{1}{4}(g_1^2+g_2^2)(3 v_1^2 - v_2^2) &
- \cos \xi \, m_{12}^2 - \tfrac{1}{2}(g_1^2+g_2^2) v_1 v_2 \\[.5em]
- \cos \xi \, m_{12}^2 - \tfrac{1}{2}(g_1^2+g_2^2) v_1 v_2 &
m_2^2 + \tfrac{1}{4}(g_1^2+g_2^2)(3 v_2^2 - v_1^2)
\end{pmatrix}, \\[.5em]
\matr{M}_{\phi \chi} &= 
\begin{pmatrix}
0 &
\sin \xi \, m_{12}^2 \\[.5em]
- \sin \xi \, m_{12}^2 &
0
\end{pmatrix}, \\[.5em]
\matr{M}_\chi &= 
\begin{pmatrix}
m_1^2 + \tfrac{1}{4}(g_1^2+g_2^2)(v_1^2 - v_2^2) &
- \cos \xi \, m_{12}^2 \\[.5em]
- \cos \xi \, m_{12}^2 &
m_2^2 + \tfrac{1}{4}(g_1^2+g_2^2)(v_2^2 - v_1^2)
\end{pmatrix}, \\[.5em]
\label{eq:massen_phipm}
\matr{M}_{\phi^\pm\phi^\pm} &= 
\begin{pmatrix}
m_1^2 + \tfrac{1}{4} g_1^2 (v_1^2 - v_2^2) + \tfrac{1}{4} g_2^2 (v_1^2
+ v_2^2) & 
- e^{i \xi} m_{12}^2 - \tfrac{1}{2} g_2^2 v_1 v_2 \\[.5em]
- e^{-i \xi} m_{12}^2 - \tfrac{1}{2} g_2^2 v_1 v_2 &
m_2^2 + \tfrac{1}{4} g_1^2 (v_2^2 - v_1^2) + \tfrac{1}{4} g_2^2 (v_1^2 + v_2^2)
\end{pmatrix}.
\end{align}
\end{subequations}
The non-vanishing elements of $\matr{M}_{\phi \chi}$ lead to
$\cp$-violating mixing terms in the Higgs potential between the $\cp$-even
fields $\phi_1$ and $\phi_2$ and the $\cp$-odd fields $\chi_1$ and
$\chi_2$ if $\xi \ne 0$. 
The mass eigenstates in lowest order follow from 
unitary transformations on the original fields, 
\begin{align}
\label{eq:RotateToMassES}
\begin{pmatrix} h \\ H \\ A \\ G \end{pmatrix} = \matr{U}_{\mathrm{n}(0)} \cdot
\begin{pmatrix} \phi_1 \\ \phi_2 \\ \chi_1 \\ \chi_2 \end{pmatrix}, \quad
\begin{pmatrix} H^\pm \\ G^\pm \end{pmatrix} = \matr{U}_{\mathrm{c}(0)} \cdot
\begin{pmatrix} \phi_1^\pm \\ \phi_2^\pm \end{pmatrix} .
\end{align}
The matrices $\matr{U}_{\mathrm{n}(0)}$ and $\matr{U}_{\mathrm{c}(0)}$ 
transform the neutral and charged Higgs fields, respectively,
such that the resulting mass matrices 
\BE
\matr{M}_{hHAG}^{\rm diag} = \matr{U}_{\mathrm{n}(0)}
\matr{M}_{\phi\phi\chi\chi} \matr{U}_{\mathrm{n}(0)}^{\dagger} 
\quad {\rm and} \quad
\matr{M}_{H^\pm G^\pm}^{\rm diag} = 
\matr{U}_{\mathrm{c}(0)} \matr{M}_{\phi^\pm\phi^\pm}
\matr{U}_{\mathrm{c}(0)}^{\dagger} 
\end{equation}
are diagonal in the basis of the transformed fields.
The new fields correspond to the three neutral Higgs bosons $h$, $H$
and $A$, the charged pair $H^\pm$ and the Goldstone bosons $G$ and
$G^\pm$. 

The lowest-order mixing matrices can be determined from
the eigenvectors of $\matr{M}_{\phi\phi\chi\chi}$ and 
$\matr{M}_{\phi^\pm\phi^\pm}$,
calculated under the additional condition that the tadpole
coefficients~(\ref{eq:phichi_tadpoles}) must vanish in order that
$v_1$ and $v_2$ are indeed stationary points of the Higgs potential. 
This automatically requires $\xi=0$, which in turn leads to a
vanishing matrix $\matr{M}_{\phi \chi}$ and a real, symmetric matrix
$\matr{M}_{\phi^\pm\phi^\pm}$. 
Therefore, no $\cp$-violation occurs in the Higgs potential at the lowest
order, and the corresponding mixing matrices can be parametrized by
real mixing angles as 
\begin{align}
\label{eq:MixMatrixO0}
  \matr{U}_{\mathrm{n}(0)} =
  \begin{pmatrix}
        - \sina & \cosa &                 0 &           0 \\
    \quad \cosa & \sina &                 0 &           0 \\
              0 &     0 &     - \sin \betan & \cos \betan \\
              0 &     0 & \quad \cos \betan & \sin \betan
  \end{pmatrix}, \quad
  \matr{U}_{\mathrm{c}(0)} =
  \begin{pmatrix}
        - \sin \betac & \cos \betac \\
    \quad \cos \betac & \sin \betac
  \end{pmatrix}.
\end{align}
The mixing angles $\alpha$, $\betan$ and $\betac$ can be determined
from the requirement that this transformation results in diagonal
mass matrices for the physical fields. 
It is necessary, however, to determine the elements of the mass matrices
without inserting the explicit form of the mixing angles and keeping the
dependence on the complex phase $\xi$, since these
expressions will be needed for the
renormalization of the Higgs potential and the calculation of the
tadpole and mass counterterms at one-loop order.


\subsection{Higgs mass terms and tadpoles}

In order to specify our notation and the conventions used in this paper
we write out the Higgs mass terms and tadpole terms in detail.
The terms in $\VHiggs$, expressed in the mass eigenstate basis,
which are linear or quadratic in the fields are denoted as follows, 
\begin{align}
\label{eq:PhysTadMass}
\VHiggs &= \mathrm{const.} - \tadh \cdot h - \tadH \cdot H -\tadA
\cdot A - \tadG \cdot G \notag \\[.5em]
&\quad + \tfrac{1}{2}
\begin{pmatrix}
  h,H,A,G
\end{pmatrix} \cdot
\begin{pmatrix}
  \mh^2  & \mhH^2 & \mhA^2 & \mhG^2 \\
  \mhH^2 & \mH^2  & \mHA^2 & \mHG^2 \\
  \mhA^2 & \mHA^2 & \mA^2  & \mAG^2 \\
  \mhG^2 & \mHG^2 & \mAG^2 & \mG^2
\end{pmatrix} \cdot
\begin{pmatrix}
  h \\ H \\ A \\ G
\end{pmatrix} + \\[.5em]
&\quad +
\begin{pmatrix}
  H^-,G^-
\end{pmatrix} \cdot
\begin{pmatrix}
  \mHpm^2  & \mHmGp^2 \\
  \mGmHp^2 & \mGpm^2
\end{pmatrix} \cdot
\begin{pmatrix}
  H^+ \\ G^+
\end{pmatrix} + \cdots \notag.
\end{align}
Our notation for the Higgs masses in this paper is such that 
lowest-order mass parameters are written in lower case, e.g.\
$\mh^2$, while loop-corrected masses are written in upper case,
e.g.\ $\Mh^2$.

For the gauge-fixing, affecting terms involving Goldstone fields in
\refeq{eq:PhysTadMass}, we have chosen the 't~Hooft--Feynman gauge.
In the renormalization we follow the usual approach where the
gauge-fixing term receives no net contribution from the renormalization
transformations. Accordingly, the counterterms derived below arise only
from the Higgs potential and the kinetic terms of the Higgs fields but
not from the gauge-fixing term.

In order to perform the renormalization procedure in a transparent way, 
we express the parameters in
$\VHiggs$ in terms of physical parameters.
In total, $\VHiggs$ contains eight independent real parameters: $v_1$,
$v_2$, $g_1^2$, $g_2^2$, $m_1^2$, $m_2^2$, $m_{12}^2$ and $\xi$, which
can be
replaced by the parameters $\MZ$, $\MW$, $e$, $\mHpm$ (or $\mA$), $\tb$,
$T_h$, $T_H$ and $T_A$. Thereby, the coupling constants $g_1$ and $g_2$ are
replaced by the electromagnetic coupling constant $e$ and the
weak mixing angle $\theta_\mathrm{w}$ in terms of
$\cw \equiv \cos \theta_\mathrm{w} = \MW/\MZ, \sw = \sqrt{1-\cw^2}$,
\begin{align} 
\label{eq:kopplungskonstanten}
  e = g_1 \, \cw = g_2 \, \sw,
\end{align}
while the $Z$~boson mass $\MZ$ and $\tb$ substitute for $v_1$ and $v_2$:
\begin{align}
\label{eq:tanb_mz}
\MZ^2 = \tfrac{1}{2}(g_1^2 + g_2^2)(v_1^2 + v_2^2), \; \tb = \frac{v_2}{v_1}.
\end{align}
The $W$~boson mass is then given by
\BE
\MW^2 = \edz g_2^2 (v_1^2 + v_2^2).
\end{equation}
The tadpole coefficients in the mass-eigenstate basis follow from the
original ones~(\ref{eq:phichi_tadpoles}) by transforming the fields
according to \refeq{eq:RotateToMassES},
\begin{subequations}
\label{eq:tadpoles}
\begin{align}
\label{eq:tadpoleH}
\tadH &= \sqrt{2} (-m_1^2 v_1 \cosa - m_2^2 v_2 \sina + \cos \xi \,
m_{12}^2 (v_1 \sina + v_2 \cosa) \\ 
&\quad - \tfrac{1}{4}(g_1^2 + g_2^2)(v_1^2 - v_2^2)(v_1 \cosa - v_2
\sina)), \notag \\ 
\label{eq:tadpoleh}
\tadh &= \sqrt{2} (+m_1^2 v_1 \sina - m_2^2 v_2 \cosa + \cos \xi \,
m_{12}^2 (v_1 \cosa - v_2 \sina) \\ 
&\quad + \tfrac{1}{4}(g_1^2 + g_2^2)(v_1^2 - v_2^2)(v_1 \sina + v_2
\cosa)), \notag \\ 
\label{eq:tadpoleA}
\tadA &= - \sqrt{2} \sin \xi \, m_{12}^2 (v_1 \cos \betan + v_2 \sin
\betan), \\ 
\label{eq:tadpoleG} 
\tadG &= -\tan(\beta-\betan) \tadA.
\end{align}
\end{subequations}

Using \refeqs{eq:kopplungskonstanten} --
(\ref{eq:tadpoles}) the original parameters can be expressed in terms of
$e$, $\tb$, $\MZ$, $\MW$, $T_h$, $T_H$, $T_A$ and either the mass of the
neutral $A$~boson, $\mA$, or the mass of the charged Higgs boson, 
$\mHpm$ (it should be noted that 
\refeqs{eq:tadpoleH}--(\ref{eq:tadpoleG}) yield
only three independent relations because of the linear 
dependence of $\tadG$ on $\tadA$). The masses $\mA$ and $\mHpm$ are
related to the original parameters by
\begin{subequations}
\label{eq:massen_A_Hp}
\begin{align}
\label{eq:masse_A}
\mA^2 &= m_1^2 \sin^2 \betan + m_2^2 \cos^2 \betan + \sin 2 \betan
\cos \xi \, m_{12}^2 \notag \\ 
&\quad - \cos 2 \betan \tfrac{1}{4}(g_1^2 + g_2^2)(v_1^2 - v_2^2), \\
\label{eq:masse_Hp}
\mHpm^2 &= m_1^2 \sin^2 \betac + m_2^2 \cos^2 \betac + \sin 2 \betac
\cos \xi \, m_{12}^2 \notag \\ 
&\quad - \cos 2 \betac \tfrac{1}{4}(g_1^2 + g_2^2)(v_1^2 - v_2^2) +
\tfrac{1}{2} g_2^2 (v_1 \cos \betac + v_2 \sin \betac)^2~.
\end{align}
\end{subequations}
Choosing $\mA$ as the independent parameter yields the following
relations,
\begin{align}
v_1 &= \frac{\wz\,\Cb\,\sw\,\cw\,\MZ}{e} \\
v_2 &= \frac{\wz\,\Sbe\,\sw\,\cw\,\MZ}{e} \\
g_1 &= e/\cw \\
g_2 &= e/\sw \\ 
m_1^2 &=    \notag 
 -\frac{1}{2}\MZ^2 \cos(2 \be) + \mA^2 \sin^2 \be/\KL\cos^2(\be -
 \betan)\KR   \\ 
&\quad\ +\Big[ \frac{e \tadh \cos\betan}{2 \cw \sw \MZ} 
   \KL \Cb \cos\betan \Sa + \Sbe (\Ca \cos\betan + 2 \Sa \sin\betan) \KR
                         \notag \\
&\qquad\ - \frac{e \tadH \cos\betan}{2 \cw \sw \MZ}
   \KL \cos(\alpha + \be) \cos \betan + 2 \cos \alpha \sin \be \sin
   \betan \KR \Big] /\KL\cos^2(\be - \betan)\KR
  \\
m_2^2 &=   \notag 
 \frac{1}{2}\MZ^2 \cos(2 \be) + \mA^2 \cos^2 \be/\KL\cos^2(\be -
 \betan)\KR 
 \\
&\quad\ - \Big[ 
\frac{e \tadH \sin\betan}{2 \cw \sw \MZ}
   \KL  \Sa \Sbe \sin\betan + \cos \be (2 \cos \betan \sin\al-
 \cos \al \sin\betan) \KR
\notag \\ &\qquad\ + 
\frac{e \tadh \sin\betan}{2 \cw \sw \MZ} 
   \KL 2 \cos\alpha \cos\be \cos\betan +\sin(\alpha + \be) \sin \betan\KR
\Big]/\KL\cos^2(\be - \betan)\KR  \\
m_{12}^2 &= \sqrt{\KL f_m^2 + f_s^2 \KR} \\
\sin\xi &\to f_s/\sqrt{f_m^2 + f_s^2} \label{eq:sinxi}\\
\cos\xi &\to f_m/\sqrt{f_m^2 + f_s^2} ,
\end{align}
where 
\begin{eqnarray}
f_m &=& \Big[  \edz \mA^2 \SZb
         +  \frac{e \tadh}{4 \cw \sw \MZ}
           \KL \cos(\be+\al) + \Cba \cos(2 \betan) \KR \notag \\
    && {}+  \frac{e \tadH}{4 \cw \sw \MZ}
           \KL \sin(\be+\al) - \Sba \cos(2 \betan) \KR \Big]
         /\KL\cos^2(\be - \betan)\KR , \\
f_s &=& -\frac{e \tadA}{2 \sw \cw \MZ\cos(\be - \betan)} .
\label{eq:fs}
\end{eqnarray}

We now give the bilinear terms of \refeq{eq:PhysTadMass} in this basis,
expressed in terms of $\mA$
or $\mHpm$, depending on which parameter leads to more compact
expressions. For the charged Higgs sector this yields, apart from
$\mHpm$ itself,
\begin{subequations}
\label{eq:fullmass1}
\begin{align}
\label{eq:mHmGp2}
\mHmGp^2 &= - \mHpm^2 \tan(\beta-\betac) \\
&\quad - \frac{e}{2 \MZ \sw \cw} \tadH
\sin(\alpha-\betac)/\cos(\beta-\betac) \notag \\ 
&\quad - \frac{e}{2 \MZ \sw \cw} \tadh
\cos(\alpha-\betac)/\cos(\beta-\betac) \notag \\ 
&\quad - \frac{e}{2 \MZ \sw \cw} i \tadA/\cos(\beta-\betan), \notag \\
\mGmHp^2 &= (\mHmGp^2)^*, \\
\mGpm^2 &= \mHpm^2 \tan^2(\beta-\betac) \notag \\
&\quad - \frac{e}{2 \MZ \sw \cw} \tadH \cos(\alpha + \beta - 2
\betac)/\cos^2(\beta-\betac) \\ 
&\quad + \frac{e}{2 \MZ \sw \cw} \tadh \sin(\alpha + \beta - 2
\betac)/\cos^2(\beta-\betac) .
\end{align}
\end{subequations}

The neutral mass matrix is more easily parametrized by $\mA$, as can
be seen from the 2$\times$2 sub-matrix of the $A$ and $G$~bosons: 
\begin{subequations}
\begin{align}
\label{eq:mAG2}
\mAG^2 &= - \mA^2 \tan(\beta-\betan) \\
&\quad - \frac{e}{2 \MZ \sw \cw} \tadH
\sin(\alpha-\betan)/\cos(\beta-\betan) \notag \\ 
&\quad - \frac{e}{2 \MZ \sw \cw} \tadh
\cos(\alpha-\betan)/\cos(\beta-\betan), \notag \\ 
\mG^2 &= \mA^2 \tan^2(\beta-\betan) \\
&\quad - \frac{e}{2 \MZ \sw \cw} \tadH \cos(\alpha + \beta - 2
\betan)/\cos^2(\beta-\betan) \notag \\ 
&\quad + \frac{e}{2 \MZ \sw \cw} \tadh \sin(\alpha + \beta - 2
\betan)/\cos^2(\beta-\betan). 
\end{align}
\end{subequations}
The $\cp$-violating mixing terms connecting the $h$-/$H$- and the
$A$-/$G$-sector are 
\begin{subequations}
\begin{align}
\mhA^2 &= \frac{e}{2 \MZ \sw \cw} \tadA
\sin(\alpha-\betan)/\cos(\beta-\betan), \\ 
\mhG^2 &= \frac{e}{2 \MZ \sw \cw} \tadA
\cos(\alpha-\betan)/\cos(\beta-\betan), \\ 
\mHA^2 &= - \mhG^2, \\
\mHG^2 &= \frac{e}{2 \MZ \sw \cw} \tadA
\sin(\alpha-\betan)/\cos(\beta-\betan). 
\end{align}
\end{subequations}
Finally, the terms involving the $\cp$-even $h$ and $H$ bosons read:
\begin{subequations}
\label{eq:fullmass2}
\begin{align}
\label{eq:masse_h_voll}
\mh^2 &= \MZ^2 \sin^2(\alpha+\beta) \\
&\quad + \mA^2 \cos^2(\alpha-\beta)/\cos^2(\beta-\betan) \notag \\
&\quad + \frac{e}{2 \MZ \sw \cw} \tadH \cos(\alpha-\beta)
\sin^2(\alpha-\betan)/\cos^2(\beta-\betan) \notag \\ 
&\quad + \frac{e}{2 \MZ \sw \cw} \tadh \frac{1}{2}
\sin(\alpha-\betan) (\cos(2 \alpha - \beta - \betan) + 3 \cos(\beta -
\betan))/\cos^2(\beta-\betan), \notag \\ 
\label{eq:masse_hH_voll}
\mhH^2 &= - \MZ^2 \sin(\alpha+\beta) \cos(\alpha+\beta) \\
&\quad + \mA^2 \sin(\alpha-\beta)
\cos(\alpha-\beta)/\cos^2(\beta-\betan) \notag \\ 
&\quad + \frac{e}{2 \MZ \sw \cw} \tadH \sin(\alpha-\beta)
\sin^2(\alpha-\betan)/\cos^2(\beta-\betan) \notag \\ 
&\quad - \frac{e}{2 \MZ \sw \cw} \tadh \cos(\alpha-\beta)
\cos^2(\alpha-\betan)/\cos^2(\beta-\betan), \notag \\ 
\label{eq:masse_H_voll}
\mH^2 &= \MZ^2 \cos^2(\alpha+\beta) \\
&\quad + \mA^2 \sin^2(\alpha-\beta)/\cos^2(\beta-\betan) \notag \\
&\quad + \frac{e}{2 \MZ \sw \cw} \tadH \frac{1}{2}
\cos(\alpha-\betan) (\cos(2 \alpha - \beta - \betan) - 3 \cos(\beta -
\betan))/\cos^2(\beta-\betan) \notag \\ 
&\quad - \frac{e}{2 \MZ \sw \cw} \tadh \sin(\alpha-\beta)
\cos^2(\alpha-\betan)/\cos^2(\beta-\betan). \notag 
\end{align}
\end{subequations}


\subsection{Masses and mixing angles in lowest order}
\label{sec:MassenMischungenTR}

The masses and mixing angles in lowest order follow from the
minimization of the Higgs potential. As mentioned above,
this leads to the requirement that 
the tadpole coefficients $T_{\{h,H,A\}}$ and all non-diagonal entries of
the mass matrices in \refeqs{eq:fullmass1}--(\ref{eq:fullmass2}) 
must vanish (the tadpole coefficient $T_G$ vanishes automatically if 
$T_A = 0$ holds). In particular, the condition $T_A = 0$ implies that the 
complex phase $\xi$ has to vanish, see \refeqs{eq:sinxi}--(\ref{eq:fs}),
so that the Higgs sector in lowest order is $\cp$-conserving. As a
consequence, the well-known lowest-order results of the rMSSM are
recovered from \refeqs{eq:fullmass1}--(\ref{eq:fullmass2}).

It follows from \refeqs{eq:mHmGp2} and (\ref{eq:mAG2}) that
the mixing angles have to obey
\begin{align}
\label{eq:equalbetas}
  \betac = \betan = \beta .
\end{align}
The lowest-order results for the Higgs masses can in principle be
obtained from \refeqs{eq:masse_h_voll} and (\ref{eq:masse_H_voll}) after
the mixing angle $\al$ has been determined from
\refeq{eq:masse_hH_voll} by requiring that the right-hand side of the
equation vanishes. More conveniently the Higgs masses can be determined
by diagonalizing the $2 \times 2$ matrix in the
$\phi_1$--$\phi_2$ basis, which corresponds to the
entries~(\ref{eq:fullmass2}) of the matrix of the neutral Higgs bosons 
in \refeq{eq:PhysTadMass} with $\alpha$ set to zero. 
The lowest-order masses read
\label{eq:mh_mH_tree}
\begin{align}
  \{\mh^2,\mH^2\} = \frac{1}{2} \left(\mA^2+\MZ^2 \mp
  \sqrt{(\mA^2 + \MZ^2)^2 - 4 \mA^2 \MZ^2 \cos^2 2 \beta} \right). 
\end{align}
For the mixing angle $\alpha$ one obtains
\begin{align}
\label{eq:Tan2Alpha}
\al = {\rm arctan}\KKL 
  \frac{-(\mA^2 + \MZ^2) \Sbe \Cb}
       {\MZ^2 \CQb + \mA^2 \SQb - \mh^2} \KKR~, ~~
 -\frac{\pi}{2} < \al < 0~.
\end{align}
Finally, combining eqs.~(\ref{eq:massen_A_Hp})
and~(\ref{eq:equalbetas}) relates the remaining masses $\mA$ and
$\mHpm$ with each other, 
\begin{align}
\label{eq:MA0_MHp_Relation}
  \mHpm^2 = \mA^2 + \cw^2 \MZ^2 = \mA^2 + \MW^2.
\end{align}
Depending on which of the masses $\mHpm$ and $\mA$ is chosen as
independent input parameter, the other mass can be determined from
\refeq{eq:MA0_MHp_Relation}. Since the $\cp$-violating mixing in the
neutral Higgs sector implies that the $\cp$-odd $A$~boson is no longer a
mass eigenstate in higher orders, we use
the charged Higgs mass $\mHpm$ as input parameter for our analysis of
the cMSSM.


\subsection{Renormalization of the Higgs potential}
\label{sec:RenormVHiggs}

We focus here on the renormalization needed for evaluating the complete
\onel\ corrections to the Higgs-boson masses and effective couplings
(the latter corresponding to effective mixing angles) in the
cMSSM. In our numerical analysis below we will also include \twol\
corrections obtained within the FD approach, which up to now are only
known for the rMSSM~\cite{mhiggsletter,mhiggslong,mhiggsAEC}
(the renormalization of the relevant one-loop contributions is described in 
\citeres{mhiggsletter,mhiggslong,mhiggsAEC,PomssmRep}). 
Also included will be the leading resummed corrections from the
(s)bottom sector~\cite{deltamb1,deltamb2,deltamb2b}, which have been
obtained in the cMSSM.

In order to derive 
the counterterms entering the one-loop corrections to the Higgs-boson
masses and effective couplings
we renormalize the parameters appearing in the
linear and bilinear terms of the Higgs potential, %
\begin{align}
  \MZ^2 &\rightarrow \MZ^2 + \dMZsq,  & \tadh &\rightarrow \tadh + \dtadh, 
  \notag \\
  \MW^2 &\rightarrow \MW^2 + \dMWsq,  & \tadH &\rightarrow \tadH +
  \dtadH, \notag \\ 
  \matr{M}_{\phi\phi\chi\chi} &\rightarrow \matr{M}_{\phi\phi\chi\chi} +
  \delta \matr{M}_{\phi\phi\chi\chi}, & 
  \tadA &\rightarrow \tadA + \dtadA, \notag \\
  \matr{M}_{\phi^\pm\phi^\pm} &\rightarrow \matr{M}_{\phi^\pm\phi^\pm} +
  \delta \matr{M}_{\phi^\pm\phi^\pm}, & 
  \tanb &\rightarrow \tanb (1+\dtanb). 
\label{eq:PhysParamRenorm}
\end{align}
We express the counterterms in the mass-eigenstate basis of the
lowest-order Higgs fields. While the parameter $\beta$ is renormalized, 
the mixing angles $\betan$ and $\betac$ (and also $\alpha$) need not be
renormalized. In carrying out the renormalization transformations it is 
therefore necessary to distinguish $\beta$ from $\betan$ and $\betac$
(as we have done in (\ref{eq:fullmass1})--(\ref{eq:fullmass2})),
i.e.\ \refeq{eq:equalbetas} should only be applied after the
renormalization transformations.

For the counterterms arising from the mass matrices we use the definitions
\begin{align}
\delta \matr{M}_{hHAG} &= \matr{U}_{\mathrm{n}(0)}
\, \de\matr{M}_{\phi\phi\chi\chi} \matr{U}_{\mathrm{n}(0)}^{\dagger} =
  \begin{pmatrix}
    \dmhsq  & \dmhHsq & \dmhAsq & \dmhGsq \\[.5em]
    \dmhHsq & \dmHsq  & \dmHAsq & \dmHGsq \\[.5em]
    \dmhAsq & \dmHAsq & \dmAsq  & \dmAGsq \\[.5em]
    \dmhGsq & \dmHGsq & \dmAGsq & \dmGsq
  \end{pmatrix} , \\[.5em]
\delta \matr{M}_{H^\pm G^\pm} &= \matr{U}_{\mathrm{c}(0)} 
\, \de\matr{M}_{\phi^\pm\phi^\pm} \matr{U}_{\mathrm{c}(0)}^{\dagger} =
  \begin{pmatrix}
    \dmHpmsq  & \dmHmGpsq \\[.5em]
    \dmGmHpsq & \dmGpmsq 
  \end{pmatrix} .
\end{align}
It should be noted that we need only seven independent
counterterms in the Higgs sector, $\de \mHpm^2$, $\de\MZ^2$, 
$\de\MW^2$, $\de T_h$, $\de T_H$, $\de T_A$ and 
$\de\tb$. This is due to the fact that in the expressions for the mass
counterterms the renormalization of the electric charge, $\de Z_e$, 
drops out at the one-loop level.
Inserting the counterterms introduced in \refeq{eq:PhysParamRenorm}
and applying the zeroth order relations 
$T_{\{h,H,A\}} = 0$ and $\betan = \betac = \beta$ in the coefficients of
the first-order expressions yields
for the $\cp$-even part of the Higgs sector
\begin{subequations}
\label{eq:HiggsMassenCTs}
\begin{align}
\dmhsq &= \dmAsq \cos^2(\alpha-\beta) + \delta \MZ^2 \sin^2(\alpha+\beta) \\
&\quad + \frac{e}{2 \MZ \sw \cw} (\dtadH \cos(\alpha-\beta)
\sin^2(\alpha-\beta) + \dtadh \sin(\alpha-\beta)
(1+\cos^2(\alpha-\beta))) \notag \\ 
&\quad + \dtanb \sinb \cosb (\mA^2 \sin 2 (\alpha-\beta) + \MZ^2 \sin
2 (\alpha+\beta)), \notag \\ 
\dmhHsq &= \frac{1}{2} (\dmAsq \sin 2(\alpha-\beta) - \dMZsq \sin
2(\alpha+\beta)) \\ 
&\quad + \frac{e}{2 \MZ \sw \cw} (\dtadH \sin^3(\alpha-\beta) -
\dtadh \cos^3(\alpha-\beta)) \notag \\ 
&\quad - \dtanb \sinb \cosb (\mA^2 \cos 2 (\alpha-\beta) + \MZ^2 \cos
2 (\alpha+\beta)), \notag \\ 
\dmHsq &= \dmAsq \sin^2(\alpha-\beta) + \dMZsq \cos^2(\alpha+\beta) \\
&\quad - \frac{e}{2 \MZ \sw \cw} (\dtadH \cos(\alpha-\beta)
(1+\sin^2(\alpha-\beta)) + \dtadh \sin(\alpha-\beta)
\cos^2(\alpha-\beta)) \notag \\ 
&\quad - \dtanb \sinb \cosb (\mA^2 \sin 2 (\alpha-\beta) + \MZ^2 \sin
2 (\alpha+\beta)), \notag 
\end{align}
which has the same form as for the rMSSM.

For the $\cp$-odd part we obtain
\begin{align}
\dmAGsq &= \frac{e}{2 \MZ \sw \cw} (-\dtadH \sin(\alpha-\beta) -
\dtadh \cos(\alpha-\beta)) - \dtanb \mA^2 \sinb \cosb, \\ 
\dmGsq &= \frac{e}{2 \MZ \sw \cw} (-\dtadH \cos(\alpha-\beta) +
\dtadh \sin(\alpha-\beta)), 
\end{align}
which again recovers the result of the rMSSM.

For the counterterms arising from the $\cp$-violating mixing terms we
obtain
\begin{align}
\dmhAsq &= + \frac{e}{2 \MZ \sw \cw} \dtadA \sin(\alpha-\beta), \\
\dmhGsq &= + \frac{e}{2 \MZ \sw \cw} \dtadA \cos(\alpha-\beta), \\
\dmHAsq &= - \dmhGsq, \\
\dmHGsq &= \dmhAsq .
\end{align}

Finally, the counterterms arising from the mass matrix of the charged
Higgs bosons read
\begin{align}
\dmHmGpsq &= \frac{e}{2 \MZ \sw \cw} (-\dtadH \sin(\alpha-\beta) -
\dtadh \cos(\alpha-\beta) - i \, \dtadA), \\ 
&\quad - \dtanb \mHpm^2 \sinb \cosb \notag, \\
\dmGmHpsq &= (\dmHmGpsq)^*, \\
\dmGpmsq &= \frac{e}{2 \MZ \sw \cw} (-\dtadH \cos(\alpha-\beta) +
\dtadh \sin(\alpha-\beta)). 
\end{align}
\end{subequations}

As mentioned above, we use $\mHpm$ as independent input parameter. The
counterterm $\dmAsq$ in the formulas above is therefore a dependent
quantity, which has to be expressed in terms of $\dmHpmsq$ using
\begin{align}
  \dmAsq = \dmHpmsq - \dMWsq ,
\label{massct}
\end{align}
which follows from \refeq{eq:MA0_MHp_Relation}.

\bigskip
For the field renormalization, which is necessary in order to obtain
finite Higgs self-energies for arbitrary values of the external
momentum, we choose to give each Higgs doublet one
renormalization constant,
\begin{align}
\label{eq:HiggsDublettFeldren}
  \cHe \rightarrow (1 + \tfrac{1}{2} \dZ{\cHe}) \cHe, \quad
  \cHz \rightarrow (1 + \tfrac{1}{2} \dZ{\cHz}) \cHz.
\end{align}
In the mass eigenstate basis, the field renormalization matrices read
\begin{subequations}
\label{eq:higgsfeldren}
\begin{align}
  \begin{pmatrix} h \\[.5em] H \\[.5em] A \\[.5em] G \end{pmatrix} \rightarrow
  \begin{pmatrix}
    1+\tfrac{1}{2} \dZ{hh} & \tfrac{1}{2} \dZ{hH} & \tfrac{1}{2}
    \dZ{hA} & \tfrac{1}{2} \dZ{hG} \\[.5em]
    \tfrac{1}{2} \dZ{hH} & 1+\tfrac{1}{2} \dZ{HH} & \tfrac{1}{2}
    \dZ{HA} & \tfrac{1}{2} \dZ{HG} \\[.5em] 
    \tfrac{1}{2} \dZ{hA} & \tfrac{1}{2} \dZ{HA} & 1+\tfrac{1}{2}
    \dZ{AA} & \tfrac{1}{2} \dZ{AG} \\[.5em] 
    \tfrac{1}{2} \dZ{hG} & \tfrac{1}{2} \dZ{HG} & \tfrac{1}{2} \dZ{AG}
    & 1+\tfrac{1}{2} \dZ{GG} 
  \end{pmatrix} \cdot
  \begin{pmatrix} h \\[.5em] H \\[.5em] A \\[.5em] G \end{pmatrix}
\end{align}
and
\begin{align}
\label{eq:dZHpGp}
  \begin{pmatrix} H^+ \\[.5em] G^+ \end{pmatrix} &\rightarrow
  \begin{pmatrix}
    1 + \tfrac{1}{2} \dZ{H^+H^-}     &     \tfrac{1}{2} \dZ{H^- G^+} \\[.5em]
        \tfrac{1}{2} \dZ{G^- H^+} & 1 + \tfrac{1}{2} \dZ{G^+G^-}
  \end{pmatrix} \cdot
  \begin{pmatrix} H^+ \\[.5em] G^+ \end{pmatrix}, \\[.5em]
\label{eq:dZHmGm}
  \begin{pmatrix} H^- \\[.5em] G^- \end{pmatrix} &\rightarrow
  \begin{pmatrix}
    1 + \tfrac{1}{2} \dZ{H^+H^-}   &     \tfrac{1}{2} \dZ{G^- H^+} \\[.5em]
        \tfrac{1}{2} \dZ{H^- G^+} & 1 + \tfrac{1}{2} \dZ{G^+G^-}
  \end{pmatrix} \cdot
  \begin{pmatrix} H^- \\[.5em] G^- \end{pmatrix}.
\end{align}
\end{subequations}

The renormalization according to \refeq{eq:HiggsDublettFeldren}
yields the following expressions for the field
renormalization constants in \refeq{eq:higgsfeldren}:
\begin{subequations}
\label{eq:FeldrenI_H1H2}
\begin{align}
  \dZ{hh} &= \sinasq \dZ{\cHe} + \cosasq \dZ{\cHz}, \\[.2em]
  \dZ{AA} &= \sinbsq \dZ{\cHe} + \cosbsq \dZ{\cHz}, \\[.2em]
  \dZ{hH} &= \sina \cosa (\dZ{\cHz} - \dZ{\cHe}), \\[.2em]
  \dZ{AG} &= \sinb \cosb (\dZ{\cHz} - \dZ{\cHe}), \\[.2em]
  \dZ{HH} &= \cosasq \dZ{\cHe} + \sinasq \dZ{\cHz}, \\[.2em]
  \dZ{GG} &= \cosbsq \dZ{\cHe} + \sinbsq \dZ{\cHz}, \\[.2em]
  \dZ{H^- H^+} &= \sinbsq \dZ{\cHe} + \cosbsq \dZ{\cHz}, \\[.2em]
  \dZ{H^- G^+} = \dZ{G^- H^+} &= \sinb \cosb (\dZ{\cHz} - \dZ{\cHe}), \\[.2em]
  \dZ{G^- G^+} &= \cosbsq \dZ{\cHe} + \sinbsq \dZ{\cHz} ~.
\end{align}
\end{subequations}
For the field renormalization constants of the $\cp$-violating
self-energies it follows,
\begin{align}
  \dZ{hA} = \dZ{hG} = \dZ{HA} = \dZ{HG} = 0,
\label{eq:dZzero}
\end{align}
which is related to the fact that the Higgs potential is $\cp$-conserving 
in lowest order and Goldstone bosons decouple.


\subsection{Renormalization conditions}
\label{sec:rencond}

We determine the one-loop counterterms by requiring the following
renormalization conditions.
The SM gauge bosons and the charged Higgs boson are renormalized
on-shell, 
\BE
\re\hSi_{ZZ}(\MZ^2) = 0, \quad \re\hSi_{WW}(\MW^2) = 0, \quad
\re\hSi_{H^+H^-}(\MHpm^2) = 0~,
\label{eq:rencond1}
\end{equation}
where the gauge-boson self-energies are to be understood as the
transverse parts of the full self-energies. 
For the mass counterterms, \refeq{eq:rencond1} yields
\begin{align}
\label{eq:mass_osdefinition}
  \dMZsq = \re \se{ZZ}(\MZ^2), \quad \dMWsq = \re \se{WW}(\MW^2),
  \quad \dmHpmsq = \re \se{H^+ H^-}(\MHpm^2)~. 
\end{align}
It should be noted that \refeqs{eq:rencond1},
(\ref{eq:mass_osdefinition}) are strict one-loop conditions. Beyond
one-loop order we define the mass of an unstable particle according to
the real part of its complex pole, see the discussion in 
\refse{subsec:massmix} below.

The results for the self-energies can be decomposed as usual in terms of
standard scalar one-loop integrals. Because of the appearance of complex
phases, the coefficients of these loop integrals could in principle be
complex. We have explicitly verified that this is not the case, i.e.\
the complex parameters appear in the results for the self-energies only 
in combinations where the imaginary parts cancel out. As a consequence, 
the only source for imaginary parts in the results for the self-energies
are the loop integrals, as in the case of the rMSSM.

As the tadpole coefficients are required to vanish, their counterterms 
follow from 
\BE
T_{\{h,H,A\}(1)} + \de T_{\{h,H,A\}} = 0~,
\end{equation}
where $T_{\{h,H,A\}(1)}$ denote the one-loop contributions to the
respective Higgs tadpole graphs: 
\begin{align}
  \dtadh = -{\tadh}_{(1)}, \quad \dtadH = -{\tadH}_{(1)}, \quad \dtadA
  = -{\tadA}_{(1)}. 
\end{align}

Concerning the field renormalization and the renormalization of $\tb$, 
we adopt the $\DRbar$~scheme, 
\begin{subequations}
\label{eq:deltaZHiggs}
\begin{align}
  \dZ{\cHe} &= \dZ{\cHe}^{\DRbar}
       \; = \; - \KKL \re \Sip_{HH \; |\al = 0} \KKR^{\rm div}, \\[.5em]
  \dZ{\cHz} &= \dZ{\cHz}^{\DRbar} 
       \; = \; - \KKL \re \Sip_{hh \; |\al = 0} \KKR^{\rm div}, \\[.5em]
  \dtanb &= \edz \KL \dZ{\cHz} - \dZ{\cHe} \KR = \dtanb^{\DRbar} 
\end{align}
\end{subequations}
i.e.\ the renormalization constants in
\refeqs{eq:deltaZHiggs} contribute only via divergent parts. In
\refeqs{eq:deltaZHiggs} the
short-hand notation $f^{\prime}(p^2) \equiv d \, f(p^2)/(d \, p^2)$ has
been used. As default value of the renormalization scale we have chosen 
in this paper $\mu^{\drbarm} = \mt$.

The $\DRbar$~renormalization of the parameter $\tb$, which is manifestly
process-independent, is convenient since there is no obvious relation of
this parameter to a specific physical observable that would favor a
particular on-shell definition. Furthermore, the
$\DRbar$~renormalization of $\tb$ has been shown to yield stable
numerical results \cite{mhiggsf1lA,feynhiggs1.2,tanbetaren}. This
scheme is also gauge-independent at the one-loop level within
the class of $R_\xi$ gauges~\cite{tanbetaren}.

The field renormalization constants completely drop out in the determination
of the Higgs-boson masses at one-loop order. They only enter via
residual higher-order effects as a consequence of the iterative
numerical determination of the propagator poles described in 
\refse{subsec:massmix} below. The $\DRbar$~scheme for the field
renormalization constants is convenient in order to avoid the possible
occurrence of unphysical threshold effects. Higgs bosons appearing as
external particles in a physical process of course have to obey proper
on-shell conditions. This issue will be discussed in 
\refse{subsec:extHiggs}.


\section{Higgs boson masses and mixings at higher orders}
\label{sec:higherorderhiggs}

\subsection{Calculation of the renormalized self-energies}
\label{subsec:SEcalc}

At the one-loop level, the renormalized self-energies, $\hSi(p^2)$,
can now be expressed 
through the unrenormalized self-energies, $\Si(p^2)$, the field
renormalization constants and the mass counterterms. As explained above,
the counterterms arise from the Higgs potential and the kinetic terms,
while the gauge-fixing term does not yield a counterterm contribution.
The renormalization prescription
of the gauge-fixing term induces counterterm contributions in the ghost
sector, see e.g.\ \citere{mwsm} for further details.
The counterterms from the ghost sector, however, contribute to the
Higgs-boson self-energies only from two-loop order on.

The renormalized self-energies read for the $\cp$-even part,
\begin{subequations}
\label{eq:renses_higgssector}
\begin{align}
\label{renSEhh}
\ser{hh}(p^2)  &= \se{hh}(p^2) + \dZ{hh} (p^2-\mh^2) - \dmhsq, \\
\label{renSEhH}
\ser{hH}(p^2)  &= \se{hH}(p^2) + \dZ{hH}
(p^2-\tfrac{1}{2}(\mh^2+\mH^2)) - \dmhHsq, \\ 
\label{renSEHH}
\ser{HH}(p^2)  &= \se{HH}(p^2) + \dZ{HH} (p^2-\mH^2) - \dmHsq,
\end{align}
and the $\cp$-odd part,
\begin{align}
\ser{AA}(p^2)  &= \se{AA}(p^2) + \dZ{AA} (p^2 - \mA^2) - \dmAsq, \\
\ser{AG}(p^2)  &= \se{AG}(p^2) + \dZ{AG} (p^2 - \tfrac{1}{2}\mA^2) -
\dmAGsq, \\ 
\ser{GG}(p^2)  &= \se{GG}(p^2) + \dZ{GG} p^2 - \dmGsq .
\end{align}
The $\cp$-violating self-energies read (using \refeq{eq:dZzero})
\begin{align}
\ser{hA}(p^2)  &= \se{hA}(p^2) - \dmhAsq, \\ 
\ser{hG}(p^2)  &= \se{hG}(p^2) - \dmhGsq, \\ 
\ser{HA}(p^2)  &= \se{HA}(p^2) - \dmHAsq, \\ 
\ser{HG}(p^2)  &= \se{HG}(p^2) - \dmHGsq
\end{align}
while for the self-energies in the charged sector one obtains
\begin{align}
\label{renSEHp}
\ser{H^- H^+}(p^2)  &= \se{H^- H^+}(p^2) + \dZ{H^- H^+} (p^2 -
\mHpm^2) - \dmHpmsq, \\ 
\ser{H^- G^+}(p^2)  &= \se{H^- G^+}(p^2) + \dZ{H^- G^+} (p^2 -
\tfrac{1}{2} \mHpm^2) - \dmHmGpsq, \\ 
\ser{G^- H^+}(p^2)  &= \ser{H^- G^+}^*(p^2), \\
\label{renSEGp}
\ser{G^- G^+}(p^2)  &= \se{G^- G^+}(p^2) + \dZ{G^- G^+} p^2 - \dmGpmsq.
\end{align}
\end{subequations}

\setlength{\unitlength}{0.099mm}
\begin{figure}[htb!]
\begin{center}
\input{fdSEn}
\caption{Generic Feynman diagrams for the $h$, $H$, $A$, $G$ self-energies 
($f$ = \{$e$, $\mu$, $\tau$, $d$, $s$, $b$, $u$, $c$, $t$\} ).
Corresponding diagrams for the $Z$~boson self-energy are obtained by
replacing the external Higgs boson by a $Z$~boson.}
\label{fig:fdSEn}
\end{center}
\end{figure}

\setlength{\unitlength}{0.096mm}
\begin{figure}[htb!]
\begin{center}
\input{fdSEc}
\caption{Generic Feynman diagrams for the $H^\pm$, $G^\pm$ self-energies 
($l$ = \{$e$, $\mu$, $\tau$\}, 
 $d$ = \{$d$, $s$, $b$\},
 $u$ = \{$u$, $c$, $t$\} ).
Corresponding diagrams for the $W$~boson self-energy are obtained by
replacing the external Higgs boson by a $W$~boson.}
\label{fig:fdSEc}
\end{center}
\end{figure}

\setlength{\unitlength}{0.099mm}
\begin{figure}[htb!]
\begin{center}
\input{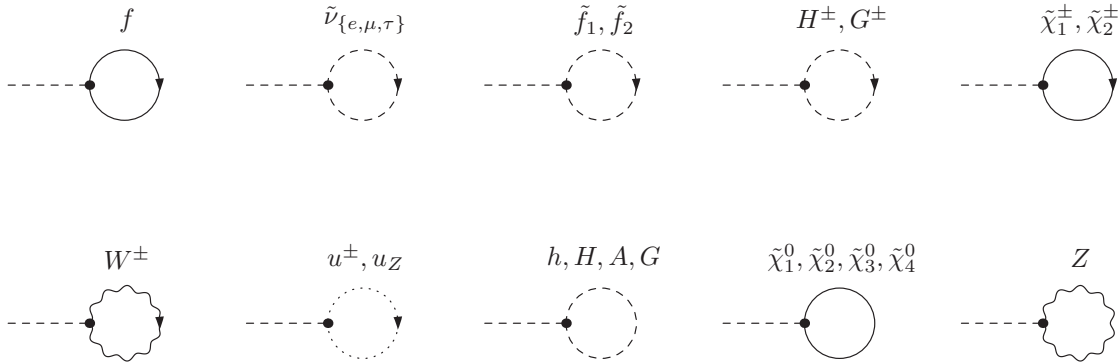}
\caption{Generic Feynman diagrams for the $h$, $H$, $A$ tadpoles
($f$ = \{$e$, $\mu$, $\tau$, $d$, $s$, $b$, $u$, $c$, $t$\}).}
\label{fig:fdTP}
\end{center}
\end{figure}

The generic Feynman diagrams for the \onel\ contribution to the Higgs
and gauge-boson self-energies are shown in \reffis{fig:fdSEn},
\ref{fig:fdSEc}. The \onel\ tadpole diagrams entering via the
renormalization are generically depicted in \reffi{fig:fdTP}. 
As usual, all the internal particles in the one-loop diagrams are
tree-level states. This implies in particular that diagrams with
internal Higgs bosons do not involve $\cp$-violating phases.
The diagrams and corresponding amplitudes have been obtained with the
program \fa~\cite{feynarts} and further evaluated with
\fc~\cite{formcalc}. As regularization scheme we have used
differential
regularization~\cite{cdr}, which has been shown to be
equivalent to 
dimensional reduction~\cite{dred} at the \onel\ level~\cite{formcalc}. 
Thus the employed regularization preserves SUSY~\cite{dredDS,dredDS2}. 

In order to obtain accurate predictions for the Higgs-boson masses and
mixings, in our numerical analysis below we will supplement the results
for the one-loop Higgs self-energies in the cMSSM obtained in this paper 
with two-loop contributions where the dependence on the complex phases
is partially neglected. The corresponding contributions will be
described in \refse{sec:beyond1l}.


\subsection{Special case: corrections to the charged Higgs-boson mass in 
the MSSM without $\cp$-violation}
\label{subsec:deltaMHp}

As a consequence of the mixing between the three neutral Higgs bosons 
in the presence of $\cp$-violating phases in the Higgs sector, it is
convenient to choose the mass of the charged Higgs boson, $\mHpm$, as
the second free input parameter in the Higgs sector besides $\tb$. In
the special case where the complex phases are zero (i.e.\ the rMSSM), 
on the other hand, one conventionally chooses the mass of the $\cp$-odd
Higgs boson, $\mA$, as independent input parameter instead of $\mHpm$, 
so that the
predictions for the neutral Higgs-boson masses do not involve charged
Higgs-boson self-energies.

In this case the mass of the charged Higgs-boson can be predicted in
terms of the other parameters and receives a shift from the higher-order
contributions. The results obtained in this paper can easily be applied
to the special case of predicting the mass of the charged Higgs boson in
the rMSSM, since the necessary ingredients are a subset of those
entering the prediction for the neutral Higgs-boson masses in the cMSSM.

The charged Higgs boson pole mass is obtained by solving
the equation
\BE
p^2 - \mHpm^2 + \hSi_{H^+H^-}(p^2) \; = \; 0~,
\end{equation}
where $\mHpm$ denotes the tree-level mass of the charged Higgs boson,
\refeq{eq:MA0_MHp_Relation}, and $\hSi_{H^+H^-}(p^2)$ is defined in
\refeq{renSEHp}. 
The mass counterterm, $\de\mHpm^2$, is given in
\refeq{massct}. In this approach, where $\mA$ is a free input parameter
($\mA = \MA$ in our notation, since the tree-level mass $\mA$ does not
receive higher-order corrections), 
the counterterm $\de\mA^2$ can be fixed by the on-shell
condition
\BE
\re\hSi_{AA}(\MA^2) = 0~,
\end{equation}
leading to
\BE
\de\mA^2 = \re\Si_{AA}(\MA^2)~.
\end{equation}
For earlier evaluations of the charged Higgs-boson mass, see
\citeres{chargedmhiggs0,chargedmhiggs}. A full \onel\ calculation
including a detailed numerical analysis can be found in \citere{markusPhD}. 


\subsection{Inclusion of higher-order corrections}
\label{sec:beyond1l}

The numerical results for the Higgs-sector observables discussed below 
are based on the complete one-loop results obtained in this paper
within the cMSSM, i.e.\ for arbitrary complex phases, supplemented by
higher-order contributions. 
The renormalized self-energies are decomposed as
\BE
\hSi(p^2) = \hSi^{(1)}(p^2) + \hSi^{(2)}(p^2) + \ldots~,
\end{equation}
where $\hSi^{(i)}$ denotes the contribution at the $i$th order. 

In addition to the full \onel\ contributions to $\hSi(p^2)$, i.e.\
$\hSi^{(1)}(p^2)$, in the cMSSM we incorporate an all-order resummation
of the $\tb$-enhanced term of \order{\alb(\als\tb)^n} including its phase 
dependence~\cite{deltamb1,deltamb2}. 
Since in the FD approach a result for the two-loop
corrections in the $t / \Stop$ sector including the full phase dependence 
is not yet available (see however \citere{mhcMSSM2L}) 
we take over the 
the leading two-loop QCD and electroweak Yukawa corrections
obtained in the
rMSSM~\cite{mhiggslong,mhiggsAEC}, neglecting the explicit phase dependence at
the two-loop level. All the contributions have been incorporated into
the Fortran code \fhtt~\cite{mhiggslong,feynhiggs,feynhiggs1.2,feynhiggs2}, see
\refse{sec:feynhiggs} below.  


\subsection{Determination of the masses from the Higgs propagators}
\label{subsec:massmix}

In order to obtain the prediction for the Higgs masses beyond lowest order, 
the poles of the Higgs propagators have to be determined. Since the
propagator poles are located in the complex plane, we define the
physical mass of each particle according to the real part of the complex
pole. 

In determining the propagator poles one needs to take into account
that the Higgs bosons mix among themselves, with the Goldstone bosons
and with the gauge bosons. For the neutral Higgs bosons of the MSSM in
the case with $\cp$ violation the Higgs propagators will in general
receive contributions from the Higgs states $h, H, A$, the Goldstone
boson $G$, and the (longitudinal part of the) $Z$~boson. The contributions 
of $G$ and $Z$ to the Higgs propagators appear from two-loop order on
via terms of the form $\left(\ser{\phi G}(p^2)\right)^2$ and 
$p^2 \left(\ser{\phi Z}(p^2)\right)^2$, where $\phi = h, H, A$ and 
$\ser{\phi Z}^\mu(p^\mu) = p^\mu \ser{\phi Z}(p^2)$. The contributions
of $G$ and $Z$ are related to each other by the usual Slavnov--Taylor
identities, ensuring a cancellation of the unphysical 
contributions. 
The mixing contributions with $G$ and $Z$ yield a sub-leading two-loop
contribution (this contribution can be compensated at the propagator poles 
by a proper choice of the field renormalization constants,
see e.g.\ \citere{susyren}). 
As explained above, we supplement the one-loop Higgs-boson
self-energies with the leading two-loop QCD and electroweak Yukawa 
corrections. Accordingly, the Higgs propagator terms induced by the mixing 
with $G$ and $Z$ are of
the same order as terms that we neglect at the two-loop level.
We will therefore neglect the effects induced by Higgs-boson mixing with 
$G$ and $Z$ in the determination of the Higgs-boson masses.%
\footnote{
We have explicitly verified that the numerical contributions
of the mixing self-energies of the Higgs bosons with $G$ and $Z$ are indeed
insignificant.
}%
~Analogously, in the charged Higgs sector we neglect the
mixing of $H^\pm$ with $G^\pm$ and $W^\pm$. While the Higgs mixing with
the Goldstone bosons and the gauge bosons yields subleading two-loop
contributions to the Higgs-boson masses, it should be noted that mixing
contributions of this kind can enter in Higgs decays or production
processes already at the one-loop level (for more details, see
\citere{karina}).

According to the discussion above we can write the propagator matrix of
the neutral Higgs bosons $h, H, A$ as a $3 \times 3$ matrix,
$\De_{hHA}(p^2)$.
(The program \fhtt\ allows to employ also the full $4 \times 4$
propagator matrix of all four scalar states $h, H, A, G$.)
The $3 \times 3$ propagator matrix
is related to the $3 \times 3$ matrix of the irreducible
vertex functions by
\begin{equation}
\De_{hHA}(p^2) = - \left(\hat{\Gamma}_{hHA}(p^2)\right)^{-1} ,
\label{eq:propagator}
\end{equation}
where 
\begin{align}
\label{eq:invprophiggs}
  \hat{\Gamma}_{hHA}(p^2) &= i \left[p^2 \unity - \matr{M}_{\mathrm{n}}(p^2)
                               \right], \\[.5em]
  \matr{M}_{\mathrm{n}}(p^2) &=
  \begin{pmatrix}
    \mh^2 - \ser{hh}(p^2) & - \ser{hH}(p^2) & - \ser{hA}(p^2) \\
    - \ser{hH}(p^2) & \mH^2 - \ser{HH}(p^2) & - \ser{HA}(p^2) \\
    - \ser{hA}(p^2) & - \ser{HA}(p^2) & \mA^2 - \ser{AA}(p^2)
  \end{pmatrix}. 
\label{eq:Mn}
\end{align}
Inversion of $\hat{\Gamma}_{hHA}(p^2)$ yields for the diagonal Higgs
propagators ($i = h, H, A$)
\begin{equation}
\De_{ii}(p^2) = \frac{i}{p^2 - m_i^2 + \ser{ii}^{\rm eff}(p^2)} ,
\label{eq:higgsprop}
\end{equation}
where $\De_{hh}(p^2)$, $\De_{HH}(p^2)$, $\De_{AA}(p^2)$ are the $(11)$,
$(22)$, $(33)$ elements of the $3 \times 3$ matrix $\De_{hHA}(p^2)$,
respectively. The structure of \refeq{eq:higgsprop} is formally the same
as for the case without mixing, but the usual self-energy is replaced by
the effective quantity $\ser{ii}^{\rm eff}(p^2)$ which contains mixing
contributions of the three Higgs bosons. It reads (no summation over 
$i, j, k$)
\begin{align}
\ser{ii}^{\rm eff}(p^2) &= \ser{ii}(p^2) - i 
\frac{2 \hat{\Gamma}_{ij}(p^2) \hat{\Gamma}_{jk}(p^2) \hat{\Gamma}_{ki}(p^2) -
      \hat{\Gamma}^2_{ki}(p^2) \hat{\Gamma}_{jj}(p^2) -
      \hat{\Gamma}^2_{ij}(p^2) \hat{\Gamma}_{kk}(p^2)
     }{\hat{\Gamma}_{jj}(p^2) \hat{\Gamma}_{kk}(p^2) - 
       \hat{\Gamma}^2_{jk}(p^2)
      } ,
\label{eq:sigmaeff}
\end{align}
where the $\hat{\Gamma}_{ij}(p^2)$ are the elements of the $3 \times 3$
matrix $\hat{\Gamma}_{hHA}(p^2)$ as specified in
\refeq{eq:invprophiggs}.

For completeness, we also state the expression for the off-diagonal 
Higgs propagators. It reads ($i \neq j$, no summation over $i, j, k$)
\begin{align}
\De_{ij}(p^2) = \frac{\hat{\Gamma}_{ij} \hat{\Gamma}_{kk} -
                      \hat{\Gamma}_{jk} \hat{\Gamma}_{ki}}{
   \hat{\Gamma}_{ii}\hat{\Gamma}_{jj}\hat{\Gamma}_{kk}
   + 2 \hat{\Gamma}_{ij}\hat{\Gamma}_{jk}\hat{\Gamma}_{ki}
   - \hat{\Gamma}_{ii}\hat{\Gamma}_{jk}^2
   - \hat{\Gamma}_{jj}\hat{\Gamma}_{ki}^2
   - \hat{\Gamma}_{kk}\hat{\Gamma}_{ij}^2
                     } ,
\label{eq:higgsprop2}
\end{align}
where we have dropped the argument $p^2$ of the $\hat{\Gamma}_{ij}(p^2)$
appearing on right-hand side for ease of notation.

The complex pole ${\cal M}^2$ of each propagator is determined as 
the solution of
\begin{equation}
{\cal M}_i^2 - m_i^2 + \ser{ii}^{\rm eff}({\cal M}_i^2) = 0 .
\end{equation}
Writing the complex pole as 
\begin{equation}
{\cal M}^2 = M^2 - i M \Ga ,
\end{equation}
where $M$ is the mass of the particle and $\Ga$ its width,
and expanding up to first order in $\Ga$ 
around $M^2$ yields the following
equation for $M_i^2$,
\begin{equation}
M_i^2 - m_i^2 + \re \ser{ii}^{\rm eff}(M_i^2) +
\frac{\im\ser{ii}^{\rm eff}(M_i^2) \, 
      \left(\im\ser{ii}^{\rm eff}\right)^\prime(M_i^2)
      }{1 + \left(\re\ser{ii}^{\rm eff}\right)^\prime(M_i^2)
       } = 0 .
\label{eq:massmaster}
\end{equation}
As before,
in \refeq{eq:massmaster} the
short-hand notation $f^{\prime}(p^2) \equiv d \, f(p^2)/(d \, p^2)$
has been used, and $M_i$ denotes the loop-corrected mass,
while $m_i$ is the lowest-order mass ($i = h, H, A$).

While the Higgs-boson masses $M_i^2$ can in principle directly be
determined from \refeq{eq:massmaster} by means of an iterative procedure 
(since $M_i^2$ appears as argument of the self-energies in
\refeq{eq:massmaster}), it is often more convenient to determine the
mass eigenvalues from a diagonalization of the mass matrix in 
\refeq{eq:Mn}. 
In our numerical analysis (and in the code \fhtt) we perform
a numerical diagonalization of \refeq{eq:Mn}
using an iterative Jacobi-type algorithm~\cite{Hahn:2006hr}. 
The mass eigenvalues $M_i$ are then determined as the zeros of the
function $\mu^2_i(p^2) - p^2$, where $\mu^2_i(p^2)$ is the $i$th eigenvalue 
of the mass matrix in \refeq{eq:Mn} evaluated at $p^2$.
Insertion of the resulting eigenvalues $M_i$ into \refeq{eq:massmaster} 
verifies (to \order{\Ga}) that each eigenvalue indeed corresponds to
the appropriate (complex pole) solution of the propagator.
We define the loop-corrected mass eigenvalues according to
\begin{equation}
M_{h_1} \leq M_{h_2} \leq M_{h_3} .
\label{eq:mh123}
\end{equation}

\medskip
In our determination of the Higgs-boson masses we take into account all
imaginary parts of the Higgs-boson self-energies 
(besides the term with imaginary parts 
appearing explicitly in \refeq{eq:massmaster}, there are also products
of imaginary parts in $\re \ser{ii}^{\rm eff}(M_i^2)$).
The effects of the imaginary parts of the Higgs-boson self-energies on
Higgs phenomenology can be 
especially relevant if the masses are close to each other. 
This has been analyzed in \citere{imagSE1} taking into account the
mixing between the two heavy neutral Higgs bosons, where the complex
mass matrix has been diagonalized with a complex mixing angle,
resulting in a non-unitary mixing matrix.
The effects of imaginary parts of the Higgs-boson self-energies on
physical processes with s-channel resonating Higgs bosons
are discussed in \citeres{imagSE1,imagSE2,imagSE3}. In
\citere{imagSE1} only the one-loop corrections from the
$t/\Stop$~sector have been taken into account for the $H$--$A$~mixing,
analyzing the effects on resonant Higgs production at a photon collider.
In \citere{imagSE2} the full one-loop imaginary parts of the
self-energies have been evaluated for the mixing of the three neutral
MSSM Higgs bosons. The effects have been analyzed for resonant Higgs
production at the LHC, the ILC and a photon collider (however, the
corresponding effects on the Higgs-boson masses have been neglected). 
In \citere{imagSE3} the $\Stop/\Sbot$ one-loop contributions (neglecting the
$t/b$ corrections) on the $H$--$A$ mixing for resonant Higgs
production at a muon collider have been discussed.
Our calculation incorporates for the first time the complete effects
arising from the imaginary parts of the one-loop self-energies in the 
neutral Higgs-boson propagator
matrix, including their effects on the Higgs masses and the Higgs
couplings in a consistent way.

As described above, the solution for the Higgs-boson masses in the
general case where the full momentum dependence and all imaginary parts
of the Higgs-boson self-energies are taken into account is numerically
quite involved. 
It is therefore of interest to consider also approximate methods 
for determining the Higgs-boson masses (often used in the literature) 
and to investigate in how far the
results obtained in this way deviate from the full result.
Instead of keeping the full momentum dependence in
\refeq{eq:Mn}, the ``$p^2$ on-shell''
approximation consists of setting the arguments of
the self-energies appearing in \refeq{eq:Mn} to the tree-level masses
according to ($i,j = h,H,A$)
\BE
\label{eq:p2onshell}
\mbox{$p^2$ on-shell approximation: } \quad
\begin{array}{lcl}
\hSi_{ii}(p^2) & \to & \hSi_{ii}(m_i^2) \\
\hSi_{ij}(p^2) & \to & \hSi_{ij}((m_i^2 + m_j^2)/2)~. \\
\end{array}
\end{equation}
In this way the Higgs-boson masses can simply be obtained as the
eigenvalues of the (momentum-independent) matrix of \refeq{eq:Mn}. 
The ``$p^2$ on-shell'' approximation has the benefit that it 
removes all residual dependencies on the field renormalization 
constants that cannot be avoided in an iterative procedure for
determining the mass eigenvalues, see the
discussion in \refse{sec:rencond}. 

Instead of setting the momentum argument of the renormalized
self-energies to the tree-level masses, in the ``$p^2 = 0$''
approximation the momentum dependence of the self-energies is neglected
completely ($i,j = h,H,A$),
\BE
\label{eq:p20approx}
\mbox{$p^2 = 0$ approximation: } \quad
\begin{array}{lcl}
\hSi_{ii}(p^2) & \to & \hSi_{ii}(0) \\
\hSi_{ij}(p^2) & \to & \hSi_{ij}(0)~. \\
\end{array}
\end{equation}
In the ``$p^2 = 0$'' approximation the masses are identified with 
the eigenvalues of $\matr{M}_{\mathrm{n}}(0)$ (see \refeq{eq:Mn})
instead of the true pole masses. This approximation is
mainly useful for comparisons with effective-potential calculations
and the determination of effective couplings (see below).
The matrix $\matr{M}_{\mathrm{n}}(0)$ is hermitian (and real and symmetric)
by construction.

In order to study the impact of the imaginary parts of the Higgs-boson
self-energies, it is useful to compare the full result with the 
``$\im\Si = 0$'' approximation, which is defined by performing 
the replacement
\begin{equation}
\mbox{$\im\Si = 0$ approximation: } \quad
\begin{array}{lcl}
\Si(p^2) & \to & \re \Si(p^2) 
\end{array}
\label{eq:ImSi0approx}
\end{equation}
for all Higgs-boson self-energies. Also this approximation results in
an hermitian mass matrix.
The comparison of our full result 
with the ``$p^2$ on-shell'', the ``$p^2 = 0$'' and the ``$\im\Si = 0$''
approximations will be discussed in \refse{sec:numeval}.


\subsection{Amplitudes with external Higgs bosons}
\label{subsec:extHiggs}

In evaluating processes with external (on-shell) Higgs bosons beyond
lowest order one has to account for the mixing between the Higgs
bosons in order to ensure that the outgoing particle has the correct
on-shell properties such that the S~matrix is properly normalized. 
This gives rise to finite wave-function normalization factors.%
\footnote{The introduction of these factors can in principle be avoided 
by using a renormalization scheme where all involved particles obey on-shell 
conditions from the start, but it is often more convenient to work in a
different scheme like the \drbar\ scheme for the field renormalizations
described in \refse{sec:RenormVHiggs}.}
For the case of $2 \times 2$ mixing appearing in the rMSSM for the
mixing between the two neutral $\cp$-even Higgs bosons $h$ and $H$, 
which is analogous to the mixing of the
photon and $Z$ boson in the Standard Model, the relevant wave function
normalization factors are well-known, see e.g.\
\citeres{mhiggsf1lC,eennH}. An amplitude with an external Higgs boson,
$i$,  receives the corrections ($i,j = h,H$, no summation over $i,j$)
\BE
\sqrt{\hat Z_i} \KL \Ga_i \; + \; \hat Z_{ij} \Ga_j \KR
                                                   \quad (i\neq j)~,
\label{eq:Vert_i}
\end{equation}
where the
$\Ga_{i,j}$ denote the one-particle irreducible Higgs vertices, and 
\begin{align}
\label{eq:Zi}
\hat Z_i &= \KKL 1 + \re \hSi_{ii}^{\prime}(p^2) - 
     \re \KL \frac{\KL \hSi_{ij}(p^2) \KR^2}
     {p^2 - m_j^2 + \hSi_{jj}(p^2)} 
     \KR^{\prime}~\KKR^{-1}_{\Bigr| p^2 = M_i^2}\,, 
                                                                 \\[.5em]
\label{eq:Zij}
\hat Z_{ij} &= -\frac{\hSi_{ij}(M_i^2)}{M_i^2 - m_j^2 + \hSi_{jj}(M_i^2)}\, .
\end{align}
As before
$m_j$ denotes the tree-level mass, while $M_i$ is the
loop-corrected mass.

In the case of the cMSSM, the formulas above need to be extended to the
case of $3 \times 3$ mixing. This can easily be achieved using the
results of \refse{subsec:massmix}. A vertex with an external Higgs
boson, $i$,
has the form (with $i,j,k$ all different, $i,j,k = h, H, A$, and no
summation over indices)
\BE
\sqrt{\hat Z_i} \KL \Ga_i \; + \; 
          \hat Z_{ij} \Ga_j \; + \; \hat Z_{ik} \Ga_k + \ldots \KR ~,
\label{eq:zfactors}
\end{equation}
where the ellipsis represents contributions from the mixing with the
Goldstone boson and the $Z$~boson, as discussed above.
The finite $Z$~factors are given by
\begin{align}
\hat Z_i &= \frac{1}{1 + 
              \left(\re\ser{ii}^{\rm eff}\right)^\prime(M_i^2)} , 
              \label{eq:zi} \\[.5em]
\hat Z_{ij} &= \frac{\De_{ij}(p^2)}{\De_{ii}(p^2)}_{~\Bigr| p^2 = M_i^2} \non \\
            &= \frac{\ser{ij}(M_i^2) 
               \left(M_i^2 - m_k^2 + \ser{kk}(M_i^2)\right) -
               \ser{jk}(M_i^2)\ser{ki}(M_i^2)
                    }{
               \ser{jk}^2(M_i^2) - 
               \left(M_i^2 - m_j^2 + \ser{jj}(M_i^2)\right)
               \left(M_i^2 - m_k^2 + \ser{kk}(M_i^2)\right)
                     } ,
               \label{eq:zij}
\end{align}
where the propagators $\De_{ii}(p^2)$, $\De_{ij}(p^2)$ 
have been given in \refeqs{eq:higgsprop} and (\ref{eq:higgsprop2}),
respectively. Using
\refeq{eq:zfactors} with the $Z$ factors specified in \refeqs{eq:zi},
(\ref{eq:zij}) and adding to this expression the mixing contributions of
the Higgs bosons with the Goldstone bosons and the gauge bosons (see the
discussion above) 
yields the correct normalization of the outgoing Higgs
bosons in the S~matrix.

For later convenience we define a matrix $\matr{\tilde Z}_{\mathrm{n}}$
based on the wave function normalization factors. Its elements are given
by (with $\hat Z_{ii} = 1$, $i, j = h, H, A$, and no summation over $i$)
\BE
(\matr{\tilde Z}_{\mathrm{n}})_{ij} := \sqrt{\hat Z_i} \; \hat Z_{ij}~.
\label{eq:defZ}
\end{equation}
Some care is necessary in order to correctly identify the elements 
$(\matr{\tilde Z}_{\mathrm{n}})_{ij}$ (given in terms of the $h, H, A$
states) with the corresponding mass eigenstates $h_1, h_2, h_3$.
To find the correct assignment, besides using \refeq{eq:massmaster} as
described above, for mass-degenerate cases we also 
compute the matrix $\matr{\tilde Z}_{\mathrm{n}}$ 
for all possible permutations of the Higgs bosons involved in the mixing 
and choose the permutation which minimizes 
$\sum_{ij} |(\matr{\tilde Z}_{\mathrm{n}})_{ij} - C_{ij}|$.
Here $C_{ij}$ is the (in general non-unitary) mixing matrix resulting
from diagonalizing the full mass matrix.%
\footnote{The matrix $C$ depends of course on the external momentum
$p^2$ where it is evaluated. Since the dependence on $p^2$ is not
very pronounced and we need $C$ only to distinguish
mass-degenerate cases, we choose the $C$ evaluated at
$p^2 = M_{h_2}^2$ since the mass ordering ensures that
the second-lightest Higgs boson is always involved in the
degeneracy.}
This procedure results in the matrix $\matr{Z}_{\mathrm{n}}$ that is
obtained from the matrix $\matr{\tilde  Z}_{\mathrm{n}}$ by a re-ordering 
of its rows.
A vertex with an external Higgs boson $h_i$ is then given by
\BE
(\matr{Z}_{\mathrm{n}})_{i1} \Ga_h +
(\matr{Z}_{\mathrm{n}})_{i2} \Ga_H +
(\matr{Z}_{\mathrm{n}})_{i3} \Ga_A + \ldots ~,
\label{eq:zfactors123}
\end{equation}
where the ellipsis again represents contributions from the mixing with the
Goldstone boson and the $Z$~boson.


\subsection{Effective couplings}
\label{subsec:effcoupl}

In a general amplitude with internal Higgs bosons, the structure describing
the Higgs part is given by
\BE
\sum_{ij} \Ga_i \; \De_{ij} \; \Ga_j
\end{equation}
where the $\Ga_{i,j}$ are as above the one-particle irreducible Higgs
vertices, and the propagators $\De_{ij}$ are given in \refeqs{eq:higgsprop}
and (\ref{eq:higgsprop2}). 
For phenomenological analyses it is often convenient to use approximations of
improved-Born type with effective couplings incorporating leading
higher-order effects. There is no unique prescription how to define such
effective coupling terms. One possibility would be to consider the matrix
$\matr{Z}_{\mathrm{n}}$, defined through 
\refeqs{eq:defZ}--(\ref{eq:zfactors123}), as mixing matrix. 
The elements of the matrix $\matr{Z}_{\mathrm{n}}$, however, are in general
complex, so that $\matr{Z}_{\mathrm{n}}$ is a non-unitary
matrix. Therefore it cannot be interpreted as a rotation matrix. 
If one wants to introduce effective couplings by means of a (unitary) rotation
matrix, it is necessary to make further approximations.

A possible choice leading to a unitary rotation matrix is the ``$p^2 = 0$''
approximation, which is used in the effective potential approach.
As before, we first consider the case of $2 \times 2$ mixing relevant
for the rMSSM. 
In the ``$p^2 = 0$'' approximation defined in \refeq{eq:p20approx}
the momentum dependence in the renormalized self-energies is neglected, 
$\hSi(p^2) \to \hSi(0)$, so that the derivative in \refeq{eq:Zi} acts
only on the $p^2$~term in the propagator factor. In this limit 
$\hat Z_i$ simplifies to~\cite{eehZhA,hff}
\BE
\mbox{$p^2 = 0$ approximation, $2 \times 2$ mixing: } \quad
\hat Z_i = \ed{1 + \hat Z_{ij}^2}~.
\label{eq:zi22p20}
\end{equation}
For the mixing between the neutral $\cp$-even Higgs bosons $h, H$ this
yields $\hat Z_h = \hat Z_H = \cos^2\De\al$. This corresponds to an
effective coupling approximation where the 
tree-level mixing angle $\alpha$
appearing in the couplings of the neutral $\cp$-even Higgs bosons is
replaced by $\aeff = \al + \De\al$~\cite{eehZhA,hff}.

It is easy to verify that for the $3 \times 3$ mixing case \refeq{eq:zi} 
in the ``$p^2 = 0$'' approximation simplifies to 
\BE
\mbox{$p^2 = 0$ approximation, $3 \times 3$ mixing: } \quad
\hat Z_i = \ed{1 + \hat Z_{ij}^2 + \hat Z_{ik}^2} ~,
\label{eq:zi33p20}
\end{equation}
as a direct generalization of \refeq{eq:zi22p20}.

The matrix $\matr{Z}_{\mathrm{n}}$ defined through 
\refeqs{eq:defZ}--(\ref{eq:zfactors123})
goes over into a unitary rotation matrix $\matr{R}_{\mathrm{n}}$ in this
approximation, 
\BE
\mbox{$p^2 = 0$ approximation, $3 \times 3$ mixing: } \quad
\matr{Z}_{\mathrm{n}} \to \matr{R}_{\mathrm{n}}, \quad
\matr{R}_{\mathrm{n}} =
  \begin{pmatrix}
    R_{11} & R_{12} & R_{13} \\
    R_{21} & R_{22} & R_{23} \\
    R_{31} & R_{32} & R_{33}
  \end{pmatrix} .
\label{eq:Rmatrix1}
\end{equation}
The matrix $\matr{R}_{\mathrm{n}}$ diagonalizes the matrix 
$\matr{M}_{\mathrm{n}}(0)$ arising from \refeq{eq:Mn} in the 
``$p^2 = 0$'' approximation. 
$\matr{R}_{\mathrm{n}}$ can therefore be used to connect 
the mass eigenstates $h_1, h_2, h_3$
with the original states $h, H, A$,
\begin{align}
  \begin{pmatrix} h_1 \\ h_2 \\ h_3 \end{pmatrix}_{p^2=0} &\hspace{-0.5em}=
  \matr{R}_{\mathrm{n}} \cdot \begin{pmatrix} h \\ H \\ A
  \end{pmatrix}, \quad 
  \matr{R}_{\mathrm{n}} \, \matr{M}_{\mathrm{n}}(0) \,
  \matr{R}_{\mathrm{n}}^\dagger =
  \begin{pmatrix}
    M_{\He,p^2=0}^2 & 0 & 0 \\ 
    0 & M_{\Hz,p^2=0}^2 & 0 \\ 
    0 & 0 & M_{\Hd,p^2=0}^2
  \end{pmatrix} .
\label{eq:defR}
\end{align}

\smallskip
We will discuss in this paper also the possibility of defining the
effective couplings in the ``$p^2$ on-shell'' approximation. 
The unitary matrix $\matr{U}_{\mathrm{n}}$ is then defined such that it
diagonalizes the matrix  
$\re\KL\matr{M}_{\mathrm{n}}(p^2 \mbox{ on-shell})\KR$ 
arising from \refeq{eq:Mn} in the 
``$p^2$ on-shell'' approximation and restricting to the real part of the
matrix.
This yields
\begin{align}
  &\mbox{$p^2$~on-shell approx., $3 \times 3$ mixing: } \;
  \begin{pmatrix} h_1 \\ h_2 \\ h_3 \end{pmatrix}_{p^2 {\rm ~on-shell}} 
  \hspace{-0.5em}=
  \matr{U}_{\mathrm{n}} \cdot \begin{pmatrix} h \\ H \\ A
  \end{pmatrix}, \;
\matr{U}_{\mathrm{n}} =
  \begin{pmatrix}
    U_{11} & U_{12} & U_{13} \\
    U_{21} & U_{22} & U_{23} \\
    U_{31} & U_{32} & U_{33}
  \end{pmatrix} , \non \\[.5em]
  &\matr{U}_{\mathrm{n}} \, \re\KL\matr{M}_{\mathrm{n}}(p^2 \mbox{ on-shell})\KR \,
  \matr{U}_{\mathrm{n}}^\dagger =
  \begin{pmatrix}
    M_{\He,p^2 {\rm ~on-shell}}^2 & 0 & 0 \\ 
    0 & M_{\Hz,p^2 {\rm ~on-shell}}^2 & 0 \\ 
    0 & 0 & M_{\Hd,p^2 {\rm ~on-shell}}^2
  \end{pmatrix} .
\label{eq:defU}
\end{align}

The elements of $\matr{U}_{\mathrm{n}}$,
which can be chosen to be real, can be used to quantify the
extent of $\cp$-violation. (The same applies to
$\matr{R}_{\mathrm{n}}$, which is real by construction.)
For example, $U_{13}^2$ can be understood as the $\cp$-odd part in $h_1$,
while the combination $U_{11}^2+U_{12}^2$ corresponds to
the $\cp$-even part. 
The unitarity of $\matr{U}_{\mathrm{n}}$ ensures that both parts
add up to 1.

The elements of $\matr{U}_{\mathrm{n}}$ (or $\matr{R}_{\mathrm{n}}$) can
be interpreted as effective couplings of Higgs bosons,
which take into account leading higher-order
contributions.
As an example, we discuss here the effective couplings of the neutral MSSM
Higgs bosons to SM gauge bosons and fermions.

Beyond the lowest order in the cMSSM all three neutral Higgs bosons have
a $\cp$-even component, so that all three Higgs bosons have
non-vanishing couplings to two gauge bosons, $VV = ZZ, W^+W^-$. 
The couplings normalized
to the SM values are given by
\begin{align}
g_{h_i VV} &= U_{i 1} \Sba + U_{i 2} \Cba .
\label{eq:hiVV}
\end{align}
The coupling of two Higgs bosons to a $Z$~boson, normalized to
the SM value, is given by
\begin{align}
g_{h_ih_jZ} &= U_{i 3} \KL U_{j 1} \Cba - U_{j 2} \Sba \KR \notag \\
            &- U_{j 3} \KL U_{i 1} \Cba - U_{i 2} \Sba \KR .
\end{align}
The Bose symmetry forbidding any anti-symmetric derivative coupling
of a vector particle to two identical real scalar fields is respected,
$g_{h_ih_iV} = 0$.

Concerning the decay into light SM fermions, 
we will compare in \refse{sec:numeval} below the full result based on
the wave function normalization factors with the effective coupling
approximation. In the latter approximation,
the decay width of $h_i$ can be obtained from the SM
decay width of the Higgs boson by multiplying it with
\BE
\KKL \KL g_{h_iff}^\rmS \KR^2 + \KL g_{h_iff}^\rmP \KR^2 \KKR,
\label{eq:hiffU}
\end{equation}
where
\begin{align}
 g_{h_iuu}^\rmS &= (U_{i 1}\Ca + U_{i 2} \Sa)/\sbe, \quad
& g_{h_iuu}^\rmP = U_{i 3} \; \cbe/\sbe \\
 g_{h_idd}^\rmS &= (-U_{i 1} \Sa + U_{i 2} \Ca)/\cbe, \quad
& g_{h_idd}^\rmP = U_{i 3} \; \sbe/\cbe
\end{align}
for up- and down-type quarks, respectively.

The results obtained by using effective couplings for simplified
calculations of cross sections or decay widths at fixed-order
perturbation theory are inherently less precise than those from a full
diagrammatic calculation at the same order. If effective couplings
are employed, their limitations should be kept in mind. It will be
shown below that for not too large values of $\MHp$ effective
couplings evaluated with $\matr{U}_{\mathrm{n}}$ give results closer
to the full calculation of \refeq{eq:zfactors123} for the
propagator corrections on external lines than those
evaluated with $\matr{R}_{\mathrm{n}}$. On the other hand, it can be shown
analytically that the effective couplings of the lightest Higgs boson
evaluated with $\matr{U}_{\mathrm{n}}$ do not decouple to the SM limit
for $\MHp \to \infty$. Decoupling can only be achieved employing
either the full calculation of \refeq{eq:zfactors123} or effective
couplings evaluated with $\matr{R}_{\mathrm{n}}$.


\section{Numerical analysis}
\label{sec:numeval}

Our results obtained in this paper extend the known results in the
literature in various ways. The results for the Higgs-boson masses and
couplings in the cMSSM 
available so far have been restricted to evaluations in the EP
approach~\cite{mhiggsCPXEP,mhiggsCPXsn} (at one-loop, neglecting the
momentum 
dependent effects) and to the RG improved \onel\ EP
method~\cite{mhiggsCPXRG1,mhiggsCPXRG2}. In
\citeres{mhiggsCPXEP,mhiggsCPXRG1,mhiggsCPXRG2} only corrections from
the (s)fermion sector and the gaugino sector have been taken into
account, and various non-logarithmic terms and momentum-dependent
corrections have been neglected. A calculation taking into account also 
contributions from the gauge-boson and Higgs sector has been performed
in \citere{mhiggsCPXsn}, however (besides neglecting momentum
dependent effects) using the parameter $m_{12}^2$ (see 
\refeq{eq:higgspotential}) as input.
Within the FD approach so far only the leading \onel\ $\mt^4$ corrections 
had been evaluated, using the on-shell renormalization
scheme~\cite{mhiggsCPXFD1}. Effects of imaginary parts of the one-loop
contributions to Higgs masses and couplings have mostly been neglected
in the above results. Some effects induced by products of imaginary
parts have been considered in \citeres{imagSE1,imagSE2,imagSE3}, see
the discussion in \refse{subsec:massmix}. 

Our results are based on the complete one-loop results in the cMSSM, 
taking into account the full dependence on the complex phases, the 
other MSSM parameters, and the external momentum. They involve a
consistent treatment of all imaginary parts appearing in one-loop
Higgs-boson self-energies that contribute to the Higgs-boson masses and
the wave-function normalization factors of external Higgs bosons. Our
one-loop results are supplemented by the dominant two-loop corrections
in the FD approach, as described in \refse{sec:beyond1l}. 
The higher-order
corrected Higgs-boson sector has been evaluated with the help of the
Fortran code \fhtt~\cite{mhiggslong,feynhiggs,feynhiggs1.2,feynhiggs2}, see
\refse{sec:feynhiggs} below.


\subsection{Parameters}

In the context of a detailed phenomenological analysis of the cMSSM
parameter space the existing constraints on $\cp$-violating parameters
from experimental bounds~\cite{pdg,plehnix} are of interest.
The complex phases appearing in the cMSSM are experimentally 
constrained by their
contribution to electric dipole moments of
heavy quarks~\cite{EDMDoink}, of the electron and 
the neutron (see \citeres{EDMrev2,EDMPilaftsis} and references therein), 
and of deuterium~\cite{EDMRitz}. While SM contributions enter 
only at the three-loop level, due to its
complex phases the cMSSM can contribute already at one-loop order.
Large phases in the first two generations of (s)fermions
can only be accommodated if these generations are assumed to be very
heavy~\cite{EDMheavy} or large cancellations occur~\cite{EDMmiracle},
see however the discussion in \citere{EDMrev1}. 
In the chargino and neutralino sector the three parameters $M_1$, $M_2$
and $\mu$ can be complex. However, there are only two physical complex
phases since one of the two phases of $M_1$ and $M_2$ can be rotated
away.
One finds that in particular the phase $\varphi_\mu$ is tightly
constrained (in the convention where $\varphi_{M_2} = 0$). 
The bounds on the phases of the third generation trilinear couplings,
on the other hand, are much weaker.
In order to illustrate the possible effects of complex phases 
we will show below results for $\varphi_{M_1}$ as well as $\varphi_{M_2}$
varied over the full parameter range. We will discuss the impact of the
experimental constraints where appropriate.
We treat the gluino mass parameter, which enters the observables
discussed below only from two-loop order on, as real, $M_3 \equiv \mgl$.

Our numerical analysis has been performed for the following set of
parameters (if not indicated differently):
\BEA
&& \msusy = 500 \gev, \; |\At| = |\Ab| = |\Atau| = 1000 \gev, 
   \; \non \\
&& |\mu| = 1000 \gev, \; |M_2| = 500 \gev, \; |M_1| = 250 \gev, \;
\mgl = 500 \gev, \non \\
&& \MHp = 150 \gev, \; \tb = 5, 15, \; \mudim = \mt = 171.4
                                                \gev~\mbox{\cite{newmt}}.
\label{parameters}
\EEA
We do not consider higher values of $\tb$, which in general enhance
the SUSY contributions to the electric dipole moments.

In order to evaluate the possible size of $\cp$-violating effects in the
Higgs sector in a conservative way we have chosen a relatively low
value of $\mHp$. Parts of the investigated parameter regions are
challenged by the Higgs search performed at
LEP~\cite{LEPHiggsSM,LEPHiggsMSSM}, depending in particular on the
parameters of the $\Stop$~sector. It should be noted, however, that
within the cMSSM the 
limits from the Higgs search are in general weaker than in the rMSSM,
giving rise even to situations where no experimental lower bound on 
$\MHe$ can be established at
all~\cite{LEPHiggsMSSM,cpx,CPXOPAL}. 

Our calculation at the one-loop level is completely general,
containing all complex phases. Concerning the numerical analysis, as
explained above, we restrict ourselves to low or moderate values of
$\tb$. Therefore the effects arising from the $b/\Sbot$~sector stay
relatively small. Consequently we do not study the effects of complex
phases from this sector, but focus on the phases of $\Xt$, $\At$, and of
the gaugino mass parameters $M_1$ and $M_2$.


\subsection{Predictions for the mass and couplings of the lightest Higgs
boson}
\label{subsec:fullvsapprox}

We begin with the predictions for the mass and couplings of the lightest 
neutral Higgs boson of the cMSSM, which are of particular interest in
view of the existing experimental bounds and of the prospective
high-precision measurement of the mass of a light MSSM Higgs boson at
the LHC and the ILC. We first compare our full result with the 
approximations discussed in \refse{subsec:massmix}.
Furthermore we investigate the effects of the phases of $M_2$ and
$M_1$.
We then compare the predictions for the partial decay widths of all
three neutral Higgs bosons to $\tau$~leptons based on the wave
function normalization factors defined in \refse{subsec:extHiggs}
with the effective coupling approximation.


\subsubsection{Comparison of the full result with approximations}

\begin{figure}[htb!]
\vspace{2em}
\centerline{\includegraphics{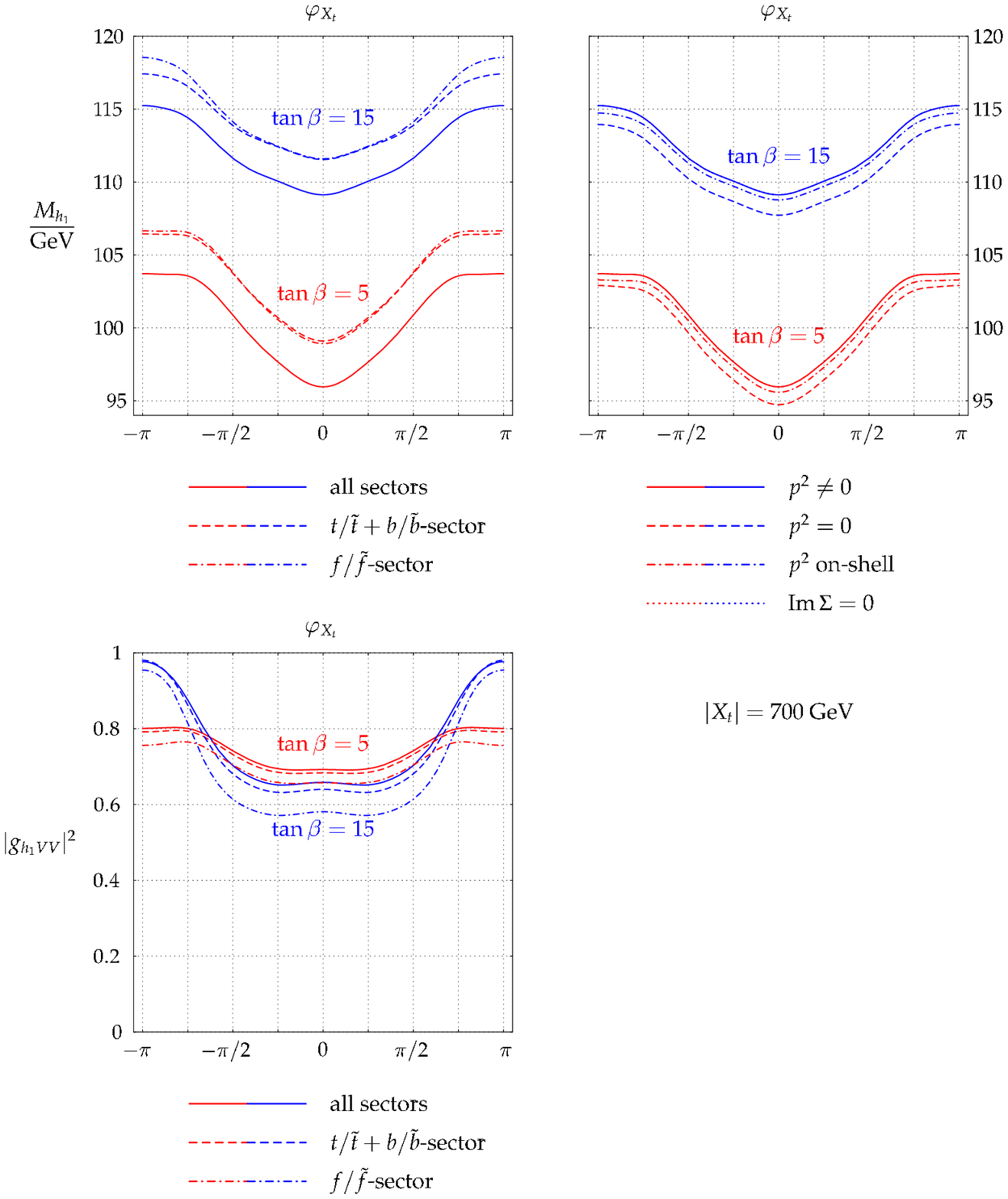}}
\caption{$\MHe$ and $|g_{\He VV}|^2$ are shown as a function of $\phixt$
  for $|\Xt| = 700 \gev$, $\tb = 5, 15$ and the
  other parameters as given in \refeq{parameters}. 
}
\label{fig:h1sferm}
\vspace{2em}
\end{figure}

In \reffi{fig:h1sferm} the cMSSM prediction for the mass of the
lightest neutral Higgs boson, $\MHe$, is shown in the upper two plots,
while the lower plot shows the coupling of $\He$ to gauge bosons. 
The results are displayed as a function of the complex phase $\phixt$
for $|\Xt| = 700\gev$. The other parameters are chosen as specified in
\refeq{parameters}. Varying $\phixt$ leaves the $\Stop$~masses
unchanged, so that the impact of the phase dependence is not masked by
the purely kinematic effect of a change in the $\Stop$~masses.
Our full result is compared with various approximations. 
In the upper left plot the full result for all sectors
of the cMSSM is compared with the results taking into account only the
effects of the $f/\Sf$~sector (dot-dashed) and from the
$t/\Stop + b/\Sbot$~sector (dashed).
It should be noted that the asymmetry between the
results for $\MHe$ at $\phixt = 0$ and $\phixt = \pm \pi$, which
amounts to about $8 \gev$ in this example, arises both from $\Xt$-dependent
one-loop corrections (whereas the leading one-loop $\mt^4$ corrections in the
limit $\MA, \MHpm \gg \MZ$ depend only on the
absolute value of $\Xt$, see e.g.\ \citere{mhiggslle}) and from two-loop
contributions. For the parameters chosen in \reffi{fig:h1sferm} there is a
partial compensation between the phase variation at the one-loop and the
two-loop contributions.
The corrections beyond the
$f/\Sf$~loops, arising from the chargino/neutralino sector, the gauge-boson
sector and the Higgs sector, can amount up to about $3 \gev$. The
$f/\Sf$~contributions are clearly dominated by the contributions of
the third generation quarks and squarks, with a maximum deviation of
about $1 \gev$ for $\phixt \approx \pm \pi$. 
Effects at the sub-GeV level may be probed at the LHC and the ILC,
where the anticipated precision for measuring the mass of a light Higgs
boson is about $0.2 \gev$ (LHC)~\cite{lhctdrs} and $0.05 \gev$ 
(ILC)~\cite{tesla,orangebook,acfarep}. For a discussion of theoretical
uncertainties from unknown higher-order corrections and the parametric
uncertainties induced by the experimental errors of the input
parameters, see e.g.\ \citeres{mhiggsAEC,PomssmRep,mhiggsWN}.

The upper right plot of \reffi{fig:h1sferm} shows the difference
between the full result and the ``$p^2$~on-shell'', 
``$p^2 = 0$'' and ``$\im\Si = 0$'' approximations defined in 
\refeqs{eq:p2onshell}--(\ref{eq:ImSi0approx}).
The ``$p^2 = 0$'' approximation yields results that differ from the full
result by up to $1.5 \gev$ in this example, while the ``$p^2$~on-shell''
approximation agrees with the full result to better than about 
$0.5 \gev$. As explained above, the imaginary parts in the one-loop 
Higgs-boson self-energies arise only from kinematical thresholds,
while the complex parameters enter only in combinations that are
real. As a consequence, for the chosen set of SUSY parameters the
self-energies entering the prediction for the lightest cMSSM Higgs
mass develop imaginary parts only from loops involving 
SM fermions (except 
the top quark). The effects of neglecting the imaginary parts are
therefore very small in this example, and the result in the 
``$\im \Si = 0$'' approximation is
indistinguishable in the plot from the full result.

The coupling of the lightest cMSSM Higgs boson to gauge bosons
normalized to the SM Higgs boson coupling, 
$|g_{\He VV}|^2$ (obtained using the ``$p^2$~on-shell'' approximation,
see \refeq{eq:hiVV}), is shown in the lower plot of
\reffi{fig:h1sferm} for the same set of parameters. This
coupling governs the Higgs production cross section in the
Higgs-strahlung channel at LEP, the Tevatron and the ILC as well as
the weak-boson fusion cross section at the LHC. 
The full result incorporating the contributions
from all sectors of the MSSM (full line) differs from the result based 
on the $f/\Sf$~sector only (dot-dashed) by 
up to 5 (10)\% in the case of $\tb = 5 \; (15)$.
The fact that the contribution from the
$t/\Stop + b/\Sbot$~sector yields a better approximation of the full
result for $|g_{\He VV}|^2$ than the contribution from the whole
$f/\Sf$~sector is due to an accidental cancellation of contributions
from different MSSM sectors.

\subsubsection{Dependence on the gaugino phases}

\begin{figure}[htb!]
\vspace{1em}
\centerline{\includegraphics{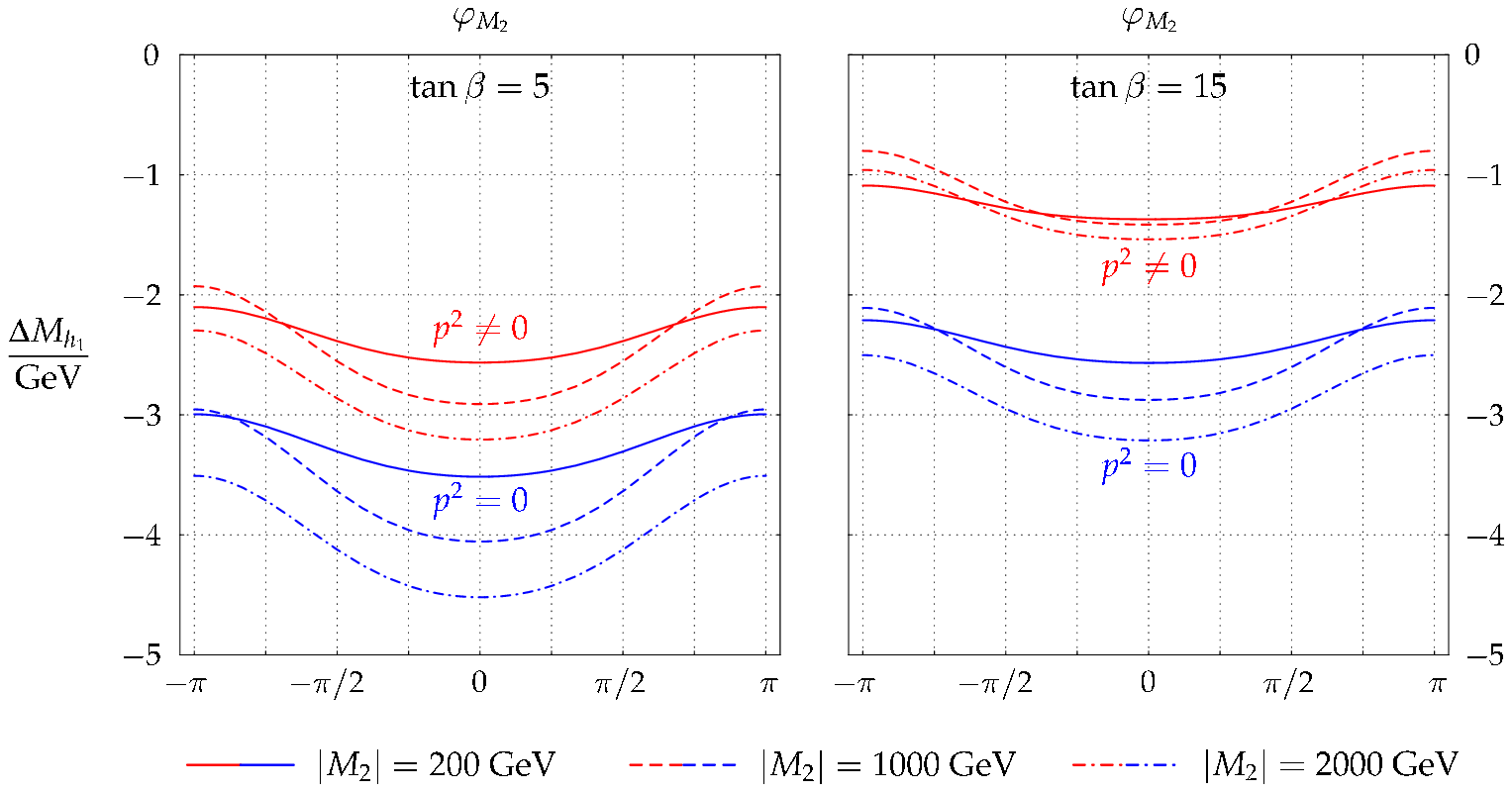}}
\caption{$\De\MHe$ := $\MHe$(all sectors) $-$ $\MHe$($f/\Sf$ sector) 
is shown as a function of
$\phiMz$ for the full result and the ``$p^2 = 0$'' approximation.
The left plot shows the result for
$\tb = 5$, while in the right plot $\tb = 15$.
$|M_2|$ is chosen as $200, 1000, 2000 \gev$.
}
\label{fig:DeMh1phiM2}
\vspace{1em}
\end{figure}

\begin{figure}[htb!]
\vspace{1em}
\centerline{\includegraphics{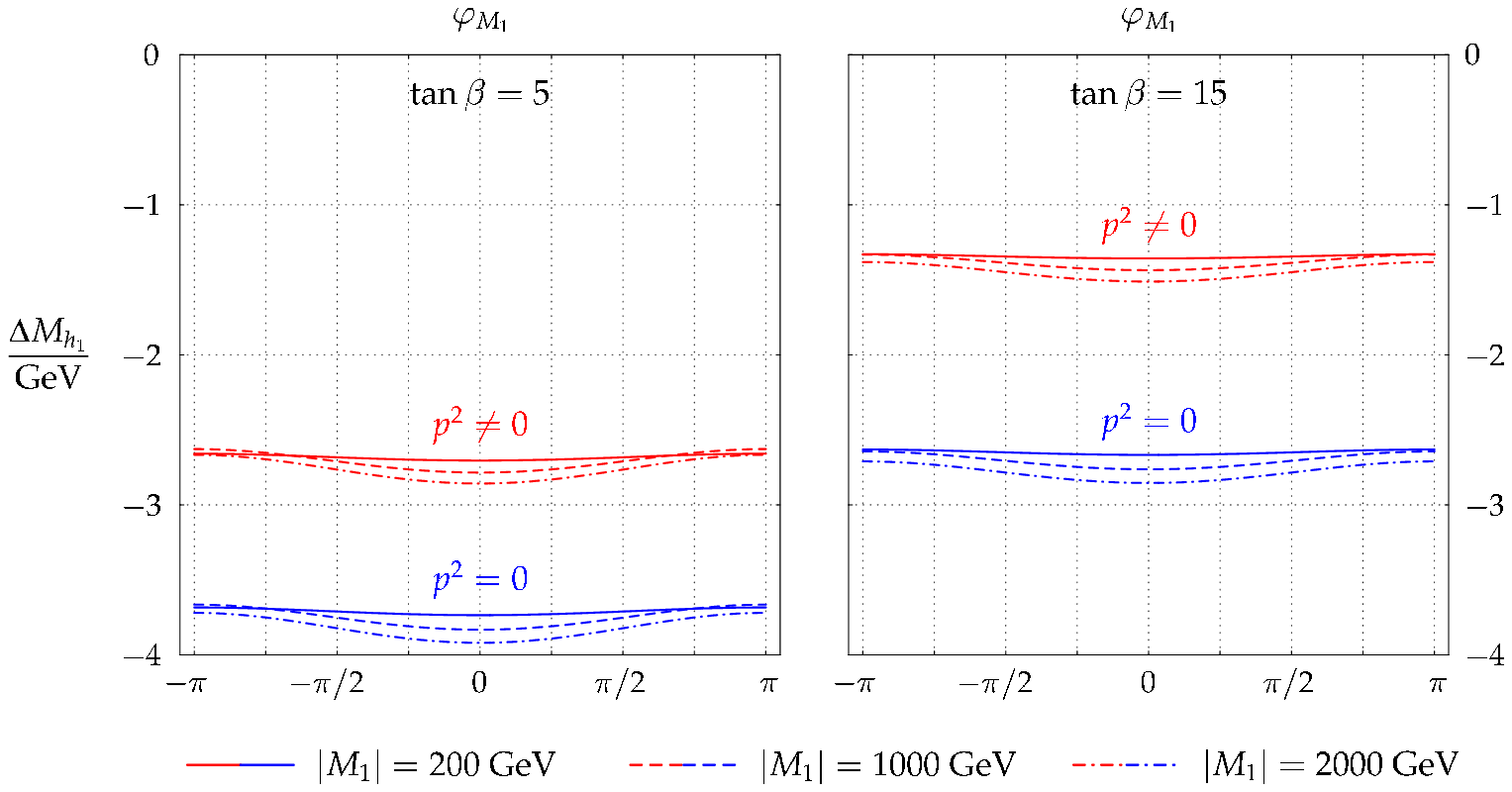}}
\caption{$\De\MHe$ := $\MHe$(all sectors) $-$ $\MHe$($f/\Sf$ sector) 
is shown as a function of
$\phiMe$ for the full result and the ``$p^2 = 0$'' approximation.
The left plot shows the result for
$\tb = 5$, while in the right plot $\tb = 15$.
$|M_1|$ is chosen as $200, 1000, 2000 \gev$.
}
\label{fig:DeMh1phiM1}
\vspace{1em}
\end{figure}

We now analyze the dependence on the gaugino phases $\phiMe$
and $\phiMz$. In \reffi{fig:DeMh1phiM2} the dependence of the
lightest cMSSM Higgs-boson mass on $\phiMz$ is shown. 
The difference
$\De\MHe$ := $\MHe$(all sectors) $-$ $\MHe$($f/\Sf$ sector),
which is dominated by
the chargino/neutralino contributions, is evaluated for three
different values of $|M_2|$, $|M_2| = 200, 1000, 2000 \gev$ (solid,
dashed, dot-dashed line). The other parameters
are chosen as in \refeq{parameters}, and all other complex phases are
set to zero. The result including the full
momentum dependence is given by the upper set of curves, while the
``$p^2 = 0$'' approximation is given by the lower set. In the left
plot we have chosen  $\tb = 5$, in the right one $\tb = 15$. 
For the lower 
$\tb$ value the effects from the non-sfermionic
sector are about $2$--$3 \gev$ if the full dependence on  
the external momentum is taken into account, and about $4 \gev$ in the
``$p^2 = 0$'' approximation (which is used in the effective potential
approach). 
The effect arising from varying the gaugino phase $\phiMz$
itself is of \order{1 \gev}. Both the overall effect from the non-sfermionic
sector and the effect from varying $\phiMz$ become smaller for
larger $\tb$ values (right plot). 
The effects are largest for $|M_2| = $\order{1 \tev}, i.e.\ for $|M_2|$ being
of the same order as $|\mu|$. In this case the gaugino-higgsino mixing in the
chargino and neutralino sector, and correspondingly the couplings of the
charginos and neutralinos to the Higgs
sector, is maximized.
The effects shown in \reffi{fig:DeMh1phiM2} 
arising from varying $\phiMz$ should be interpreted as an upper bound
on the possible impact of the phase dependence. The possible effects
from the gaugino phases will be reduced if the existing experimental
constraints on these phases are taken into account, see the discussion
above.

We now turn to the effects from varying $\phiMe$ as shown in
\reffi{fig:DeMh1phiM1}. The parameters are as in
\reffi{fig:DeMh1phiM2}, but with  $M_2 = 500 \gev$ and 
$|M_1| = 200, 1000, 2000 \gev$ (solid, dashed, dotted line).
The size of the effects from the non-sfermion sector is the same
as in \reffi{fig:DeMh1phiM2}. However, the dependence on $\phiMe$ is much
smaller, being of \order{100 \mev}.


\subsubsection{Decay widths of the neutral Higgs bosons}

\begin{figure}[htb!]
\vspace{4em}
\centerline{\includegraphics{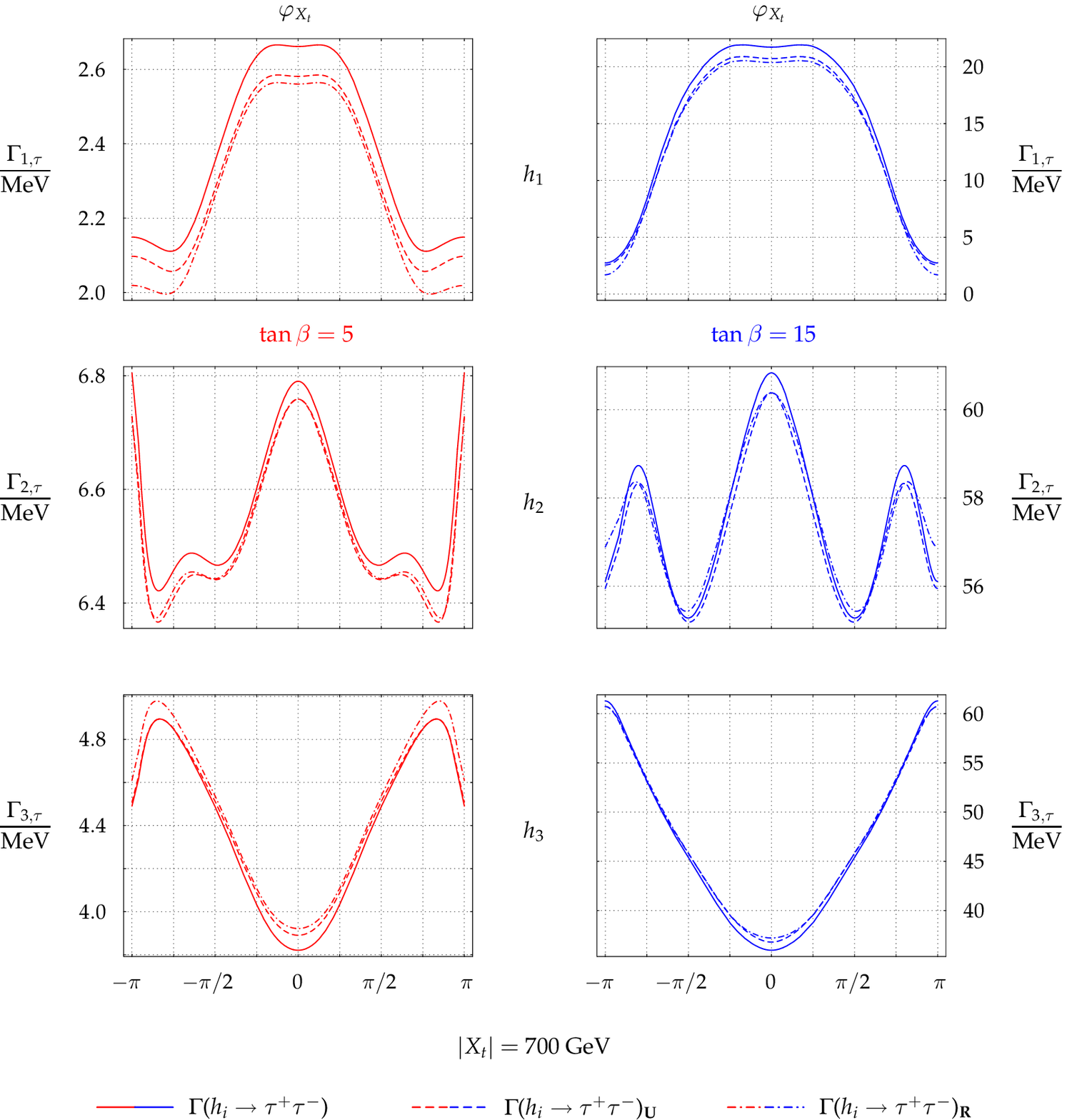}}
\caption{The decay widths
$\Ga(\Hi \to \tau^+\tau^-)$, $\Ga(\Hi \to \tau^+\tau^-)_{\matr{U}}$ 
and $\Ga(\Hi \to \tau^+\tau^-)_{\matr{R}}$ (see text)
are shown for $i = 1,2,3$ (upper, middle, lower row) as a function of
$\varphi_{\Xt}$ with $|\Xt| = 700 \gev$. In the left column $\tb = 5$,
while for the plots in the right column $\tb = 15$. The other 
parameters are chosen according to \refeq{parameters}.
}
\label{fig:h123couplings}
\vspace{2em}
\end{figure}

In this section we compare the predictions for the partial decay
widths of all three neutral Higgs bosons to $\tau$~leptons based on the wave
function normalization factors as given in 
\refeqs{eq:defZ}, (\ref{eq:zfactors123})
with the effective coupling approximation 
(using the ``$p^2$~on-shell'' approximation)
according to \refeqs{eq:defR}, (\ref{eq:defU}),
and with the ``$p^2 = 0$'' approximation as given in
\refeqs{eq:Rmatrix1}, (\ref{eq:defR}).  
In \reffi{fig:h123couplings} we show
\BE 
\Ga_{i,\tau} := \Ga(h_i \to \tau^+\tau^-) \mbox{~~and~~} 
\Ga(h_i \to \tau^+\tau^-)_{\matr{R}}, \;
\Ga(h_i \to \tau^+\tau^-)_{\matr{U}}
\end{equation}
for $i = 1, 2, 3$ (upper, middle, lower row), where $\Ga$ refers to the full
result based on the wave function normalization factors, and $\Ga_{\matr{U}}$,
$\Ga_{\matr{R}}$ correspond to the effective coupling approximation evaluated
with $\matr{U}_{\mathrm{n}}$ and $\matr{R}_{\mathrm{n}}$, respectively. 
Since we are only interested here in the comparison of the wave
function normalization factors with the effective coupling
approximations, we omit the contributions arising from the mixing of
the physical Higgs states with the Goldstone boson and the $Z$~boson 
and we also do not take into account irreducible vertex corrections to
the $h_i \tau^+\tau^-$ vertices.
The results are shown for $\MHp = 150 \gev$ and $|\Xt| = 700 \gev$ as
a function of $\phixt$, 
where the other parameters are chosen according to \refeq{parameters}
with $\tb = 5$ (left) and $\tb = 15$ (right). 
As a general feature it can be observed that $\Ga_{\matr{U}}$ is closer
to the full result 
$\Ga$ than $\Ga_{\matr{R}}$ with only few exceptions (due to
accidental numerical cancellations)%
\footnote{
For large values of $\MHp$ due to the non-decoupling effects in 
$\Ga_{\matr{U}}$, see the discussion in 
\refse{subsec:effcoupl}, $\Ga_{\matr{R}}$ would give results closer to
the full evaluation.
}%
. This shows that the effective coupling defined through
$\matr{U}_{\mathrm{n}}$, \refeq{eq:defU}, gives a somewhat better
numerical description than the 
one defined through  $\matr{R}_{\mathrm{n}}$, \refeq{eq:defR}, as used
in the effective potential approach.  
For $\tb = 5$ the deviations between the ``$p^2 = 0$'' 
approximation and the full result are mostly at or below the 5\%~level, 
where the largest effects in general occur in the decay width of the 
lightest Higgs boson. 
For $\tb = 15$ the absolute and relative deviations between the effective
coupling approximation and the full result can be significantly
larger in the case of the lightest Higgs boson. For the decay
widths of  $h_1$ the full result can differ from the 
``$p^2 = 0$'' approximation
by more than 10\%. In particular, in the region where 
$\Ga(h_1 \to \tau^+\tau^-)$ has a minimum ($\phixt \approx \pm\pi$)
the relative deviation between the full result and $\Ga_{\matr{R}}$ 
reaches more than 25\%. Also in this case the deviation between the full
result and $\Ga_{\matr{U}}$, 
based on the ``$p^2$~on-shell'' approximation, is much smaller.
The deviations for $h_2$ and $h_3$ are again at the level of 5\%.
For larger values of $|\Xt|$ even larger differences between the
effective coupling approximation and the full result can be found.

\subsection{Mass difference and mixing of the heavy neutral Higgs bosons} 
\label{subsec:heavyHiggs}

We now turn to the predictions for the masses and the mixing of the
heavy neutral Higgs bosons of the cMSSM. The discovery of heavy Higgs
bosons (in addition to a light one) would clearly establish an
enlarged Higgs sector as compared to the SM. In the cMSSM the two heavy
neutral Higgs bosons $h_2$ and $h_3$ are in general relatively close in 
mass, so that the mixing
induced by the $\cp$-violating phases can give rise to resonance-type
effects.
We first analyze the mass difference of the two
heavy neutral Higgs bosons, $\De M_{32} := \MHd - \MHz$, 
in scenarios where the Higgs-boson self-energies can be enhanced by
threshold effects. We then investigate 
the phase dependence of $\De M_{32}$ and discuss in how far this
observable can be employed for distinguishing the cMSSM from the rMSSM. 
Finally we perform a detailed analysis of the mixing of
$\Hz$ and $\Hd$ that is induced by the presence of complex phases.


\subsubsection{Threshold effects for heavy Higgs bosons}
\label{subsubsec:thresholds}

\begin{figure}[htb!]
\centerline{\includegraphics{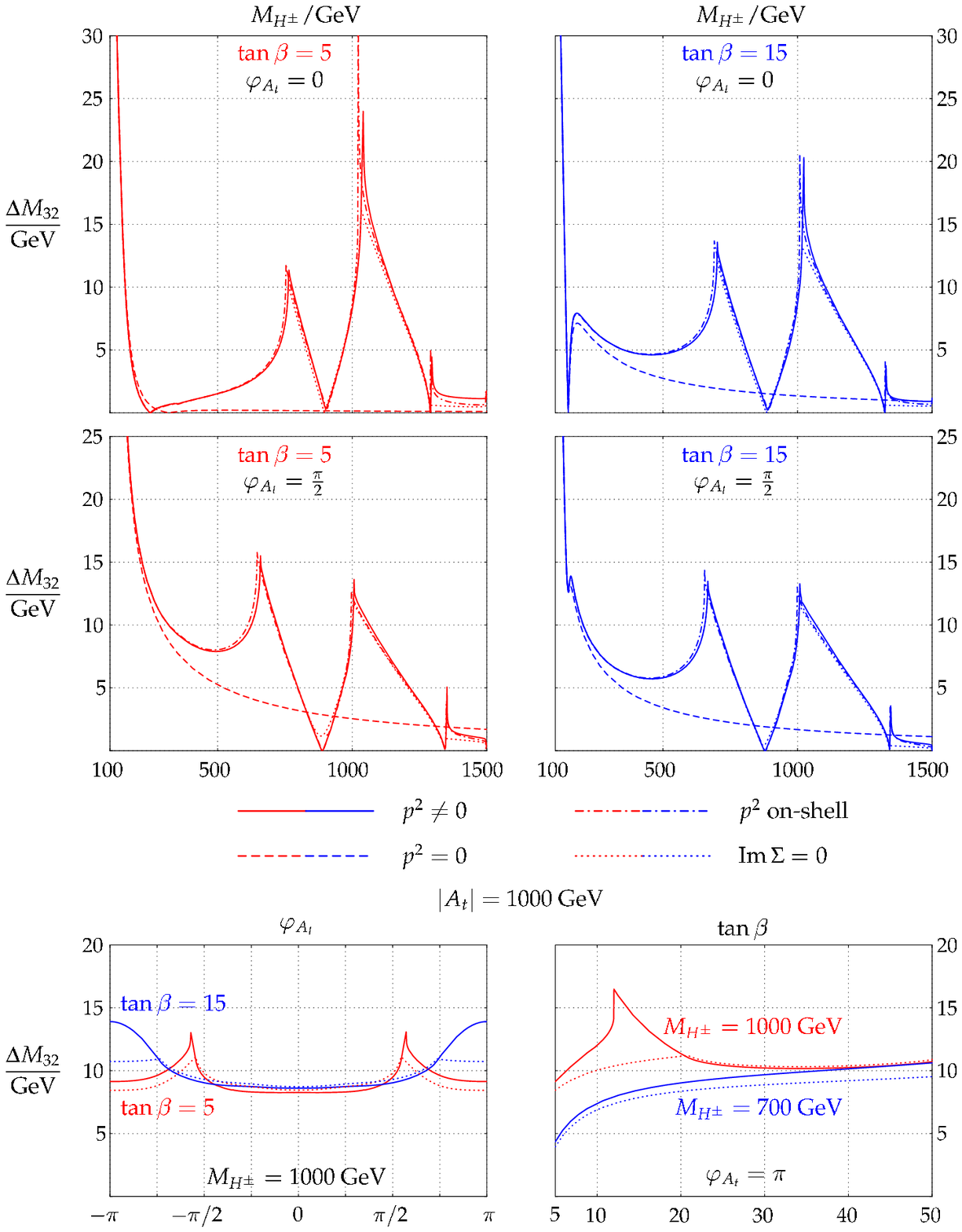}}
\caption{The mass difference $\De M_{32} := \MHd - \MHz$ is shown as a
  function of $\MHp$ (upper and middle rows) and as a function of
  $\phiat$ and $\tb$ (lower row) 
for the parameters given in \refeq{parameters}. 
The upper row shows the results for $\phiat = 0$ with $\tb = 5$
(left) and $\tb = 15$ (right). 
The middle row shows $\De M_{32}$ for $\phiat = \pi/2$ with
$\tb = 5$ (left) and $\tb = 15$ (right). 
The lower row shows $\De M_{32}$ for $\tb = 5, 15$ and $\MHp = 1000 \gev$
as a function of $\phiat$ (left) and for $\phiat = \pi$, 
$\MHp = 700, 1000 \gev$ as a function of $\tb$ (right). 
}
\label{fig:DeM32}
\end{figure}

We first
analyze the effects of thresholds appearing in the 
Higgs-boson self-energy
diagrams (e.g.\ for $\mA = \mste + \mstz$, see
the sixth diagram in \reffi{fig:fdSEn}) on the masses of
the heavy neutral Higgs bosons. In the first two lines of
\reffi{fig:DeM32} the mass difference $\De M_{32} := \MHd - \MHz$ is
shown as a function of $\mHp$ for 
$\tb = 5$ (left) and $\tb = 15$ (right) for two different values of
the phase of $\At$, $\phiat = 0$ (upper row) and $\phiat = \pi/2$
(middle row). 
The other parameters are chosen as in \refeq{parameters}. 
We compare the full result (solid lines) with the ``$p^2$~on-shell''
(dot-dashed), ``$p^2 = 0$'' (dashed)  and ``$\im\Si = 0$'' (dotted)
approximations defined in \refeqs{eq:p2onshell}--(\ref{eq:ImSi0approx}).
It can be seen for the full result that the threshold effects may lead
to a significant enhancement of the mass splitting between the states 
$h_2$ and $h_3$, so that mass differences in excess of
$\De M_{32} = 10\gev$ can occur even for $\mHp$ values in the
decoupling region where $\mHp \gg \MZ$. This behaviour is not
reproduced in the ``$p^2 = 0$'' approximation (which is used in the
effective potential approach). On the other hand, it
turns out that the ``$p^2$~on-shell'' approximation, see
\refeq{eq:p2onshell}, gives a rather good approximation to the full
result. The remaining deviations stay below the level of $1 \gev$.
It should be noted in this context that the sharp peaks displayed 
in \reffi{fig:DeM32} would get smoothened if the effects of finite 
widths of the internal particles in the Higgs-boson self-energies were
taken into account. A precise prediction directly at threshold would
require a dedicated analysis that is beyond the scope of the present
paper.

We now turn to the effects of the imaginary parts of the Higgs-boson
self-energies, i.e.\ the comparison of the full result with the 
``$\im\Si = 0$'' approximation as defined in \refeq{eq:ImSi0approx}.
While for the example of the lightest cMSSM Higgs
boson, shown in \reffi{fig:h1sferm}, the result for neglected
imaginary parts was indistinguishable from the full result, for the
mass difference of the heavy Higgs bosons a difference becomes
visible in the two upper lines of \reffi{fig:DeM32} around the
thresholds. The effect is shown in more detail in the lower line of
\reffi{fig:DeM32}. In the lower left plot we show $\De M_{32}$ as a function
of $\phiat$ for $\tb = 5, 15$ and $\mHp = 1000 \gev$. 
In the right plot we
display $\De M_{32}$ as a function of $\tb$ for $\mHp = 700, 1000 \gev$ and
$\phiat = \pi$.
The other parameters are again those of \refeq{parameters}. 
As one can see from the plot, the difference between the full result
and the approximation with neglected imaginary parts can be as large as
about $5 \gev$.


\subsubsection{Phase dependence of \boldmath{$\De M_{32}$}}
\label{subsubsec:DeM32phiAt}

\newcommand{\XAt}{\Xt}

\begin{figure}[htb!]
\centerline{\includegraphics{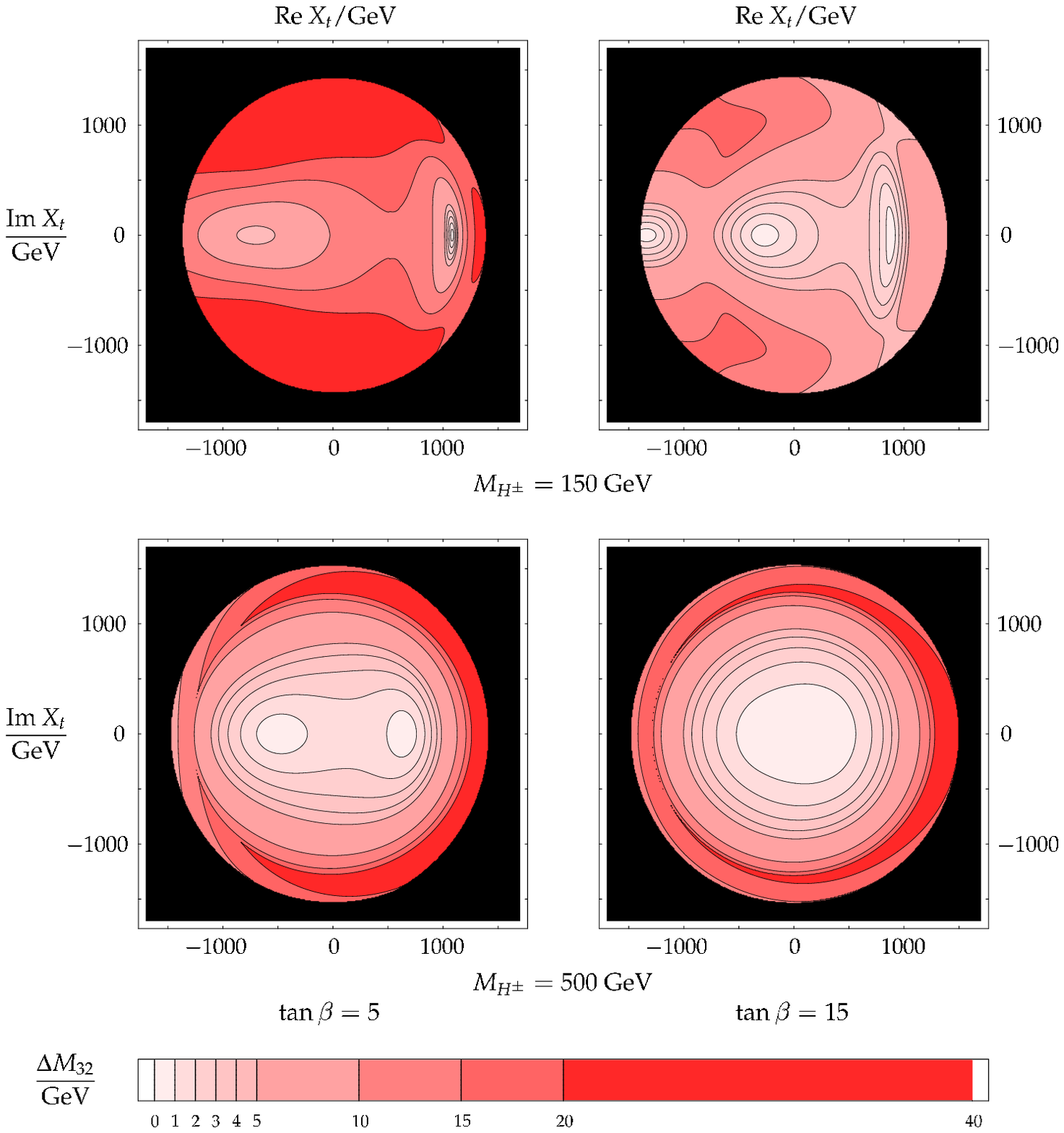}}
\caption{$\De M_{32} := \MHd - \MHz$ is shown in the $\re \Xt$--$\im \Xt$ plane 
for $\tb = 5$ (left) and $\tb = 15$ (right) for $\mHp = 150 \gev$ (upper
row) and $\mHp = 500 \gev$ (lower row).
The other parameters are as given in \refeq{parameters}.
}
\label{fig:DeM32At}
\vspace{2em}
\end{figure}

We now analyze the dependence of the mass difference 
$\De M_{32} := \MHd - \MHz$ on the complex phases in more detail, in
particular in view of the question whether the detection of a certain
mass difference between the two heavy neutral Higgs bosons could be a
direct indication of non-zero complex phases in the MSSM.
In \reffi{fig:DeM32At} we show $\De M_{32}$
in the $\re \XAt$--$\im \XAt$ plane for
$\tb = 5$ (left) and $\tb = 15$ (right) for $\mHp = 150 \gev$ (upper row) and
$\mHp = 500 \gev$ (lower row). The
other parameters are given in \refeq{parameters}.

The results in \reffi{fig:DeM32At} show that the smallest mass
differences between $\MHd$ and $\MHz$ occur if 
$\Xt \equiv \At - \mu/\tb$
is real or has only a relatively small imaginary part. 
For $\mHp = 150 \gev$ and low $\tb$ this happens only around 
$\Xt \approx 1000 \gev$, while for higher $\tb$ three minima are reached for 
$\Xt \approx -1200, -200, 800 \gev$. The largest mass differences are realized
for relatively large $|\XAt|$. While for $\tb = 5$ all possible mass
differences occurring in the $\re \XAt$--$\im \XAt$ plane 
are also realized on the real axis, for $\tb = 15$ the largest mass
differences can only be found for a non-zero imaginary part of $\XAt$.

The qualitative behaviour changes somewhat for $\mHp = 500 \gev$. 
Again the smallest mass differences between $\MHd$ and $\MHz$ occur for small
imaginary parts of $\XAt$. Two minima are found for $\tb = 5$. On the other
hand, for $\tb = 15$ only the region of the $\re\XAt$--$\im\XAt$ parameter 
space around $\XAt = 0$ results in a small value of $\De M_{32}$. The
rather symmetric shape of the plot for $\tb = 15$ and $\mHp = 500 \gev$ 
around $\XAt = 0$ shows that in this case the dominant contribution to 
$\De M_{32}$ depends only on the 
absolute value $|\Xt| \approx |\At|$. 
For $\tb = 5$, on the other hand, the minimum
values of $\De M_{32}$ are reached for both $\re \At \neq 0$ and 
$\re\Xt \neq 0$. 
Similarly to the case of $\mHp = 150 \gev$ and low $\tb$ we find also
for $\mHp = 500 \gev$ (both for low and high $\tb$) that a large mass 
difference
$\De M_{32}$ does not necessarily require a non-zero complex phase of
$\XAt$. Indeed, for large $\mHp$ all 
mass differences realized for a parameter point in the complex $\XAt$ plane
are also realized on its real axis. This means that the determination of
the mass difference
$\De M_{32}$ alone will in general not be sufficient to obtain
direct information about the size of the complex phases. 
On the other hand, the interpretation of an observed mass difference 
in terms of the underlying SUSY parameters will be different in the 
rMSSM and the cMSSM.
Valuable information for determining the parameters of the cMSSM
including their complex phases can therefore be obtained by combining
the mass difference $\De M_{32}$ with a suitable set of observables 
that exhibit a 
non-trivial dependence on the complex phases.

We have investigated also the effects of the complex phases of $M_2$ and $M_1$
on the mass difference $\De M_{32}$. The effects stay below the 
$1 \gev$~level for most parts of the parameter\,space.


\subsubsection{Mixing of the heavy neutral Higgs bosons}
\label{subsubsec:higgsmix}

We finally analyze the mixing of the two heavy neutral Higgs bosons. 
Mixing effects, especially
between the second and the third Higgs boson, can potentially be
sizable in the cMSSM. In \citere{mhiggsCPXsn} it was argued that 
the phases of the gaugino mass parameters play an important
role in this context.

As explained above, the elements of $\matr{U}_{\mathrm{n}}$ 
(or $\matr{R}_{\mathrm{n}}$) can be
used to quantify the extent of $\cp$-violation. 
For example, the combination 
$U_{31}^2 + U_{32}^2$ can be understood as the $\cp$-even part in $h_3$,
while the component $U_{33}^2$ corresponds to 
the $\cp$-odd part. 
The unitarity of $\matr{U}_{\mathrm{n}}$ ensures that both parts
add up to 1. 

\begin{figure}[htb!]
\centerline{\includegraphics{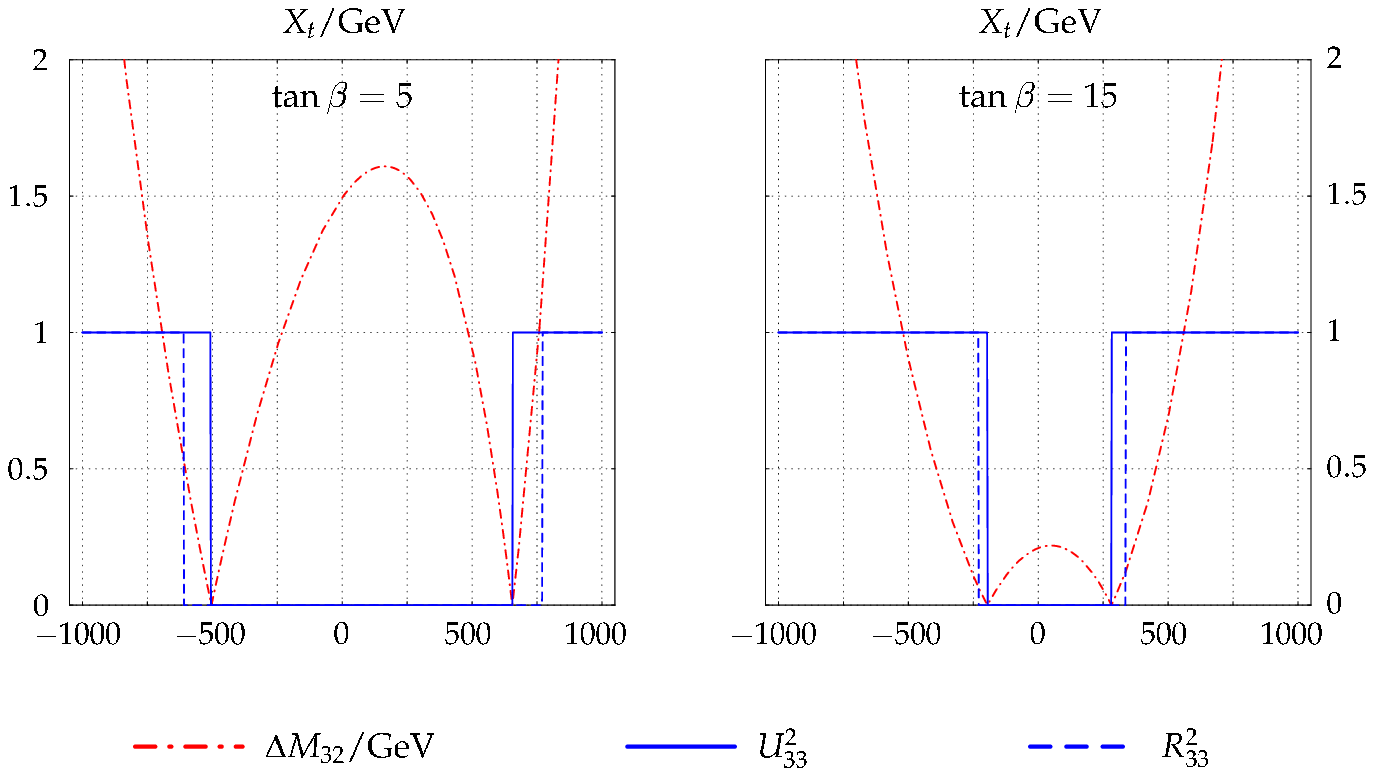}}
\caption{The mass difference $\De M_{32} = |\MH - \MA|$ and
the effective couplings $U_{33}^2$ and 
$R_{33}^2$ (based on the ``$p^2$~on-shell'' and ``$p^2 = 0$'' 
approximations, respectively)
are shown as a function of $\Xt$ (chosen
to be real) for $\MHp = 500 \gev$, $\tb = 5$ (left) and $\tb = 15$ (right).
The other parameters are chosen according to
\refeq{parameters}.
}
\label{fig:UvsR}
\end{figure}

We first discuss the $\cp$-conserving case, where $U_{33}^2$ is either~1 or~0,
depending on the mass ordering of $\MH$ and $\MA$ (the higher-order corrected
masses of the heavy neutral $\cp$ eigenstates). 
In \reffi{fig:UvsR} we show the mass difference 
$\De M_{32} = \MHd - \MHz = |\MH - \MA|$ together with the effective 
couplings $U_{33}^2$ (based on the ``$p^2$~on-shell'' approximation, see
\refeq{eq:defU}) and $R_{33}^2$ (obtained in the 
``$p^2 = 0$'' approximation, see \refeq{eq:defR}).
The three quantities are given as 
a function of $\Xt$, which in the $\cp$-conserving case is a real
parameter. We have set 
$\MHp = 500 \gev$, $\tb = 5$ ($\tb = 15$) in the left (right) plot, and the
other parameters are chosen according to 
\refeq{parameters}. 
The change from~1 to~0 in the $(33)$ element of the rotation matrix
should obviously occur at the same value of $\Xt$ where the mass hierarchy of 
the states $H$ and $A$ is inverted, i.e.\ where $\De M_{32} = 0$. This
correlation between the masses and the effective couplings is not automatic, 
however, since the masses have been calculated using the full
higher-order corrections, while as discussed in \refse{subsec:effcoupl} 
the effective couplings can only be obtained using certain
approximations. \reffi{fig:UvsR} shows that the behaviour of $U_{33}^2$
with $\Xt$ is well matched to the one of $\De M_{32}$, i.e.\ the step
in $U_{33}^2$ occurs very close to the $\Xt$ value where 
$\De M_{32} = 0$. For $R_{33}^2$, on the other hand, the behaviour of
the effective coupling significantly differs from the one of the
higher-order corrected masses. This effect is particularly pronounced
for small $\tb$ as can be seen in the left plot, where the value of $\Xt$ 
for which $\De M_{32} \approx 0$ is reached differs by more than 
$100 \gev$ from the corresponding $\Xt$ value for which
$R_{33}^2$ changes from~0 to~1. For $\tb = 15$ (right plot) the
deviation is smaller but still significant.
\reffi{fig:UvsR} clearly shows that those contributions which are omitted
if effective couplings are constructed using the ``$p^2 = 0$''
approximation (as done in the effective potential approach) can be
numerically sizable and important for a physically well-behaved result.
We find also in this case (for not too large values of $\MHp$) that a
better numerical description is obtained with effective couplings
defined through $\matr{U}_{\mathrm{n}}$, \refeq{eq:defU}. At large
$\MHp$ values both matrices 
are insufficient for a precise description.

\begin{figure}[htb!]
\centerline{\includegraphics{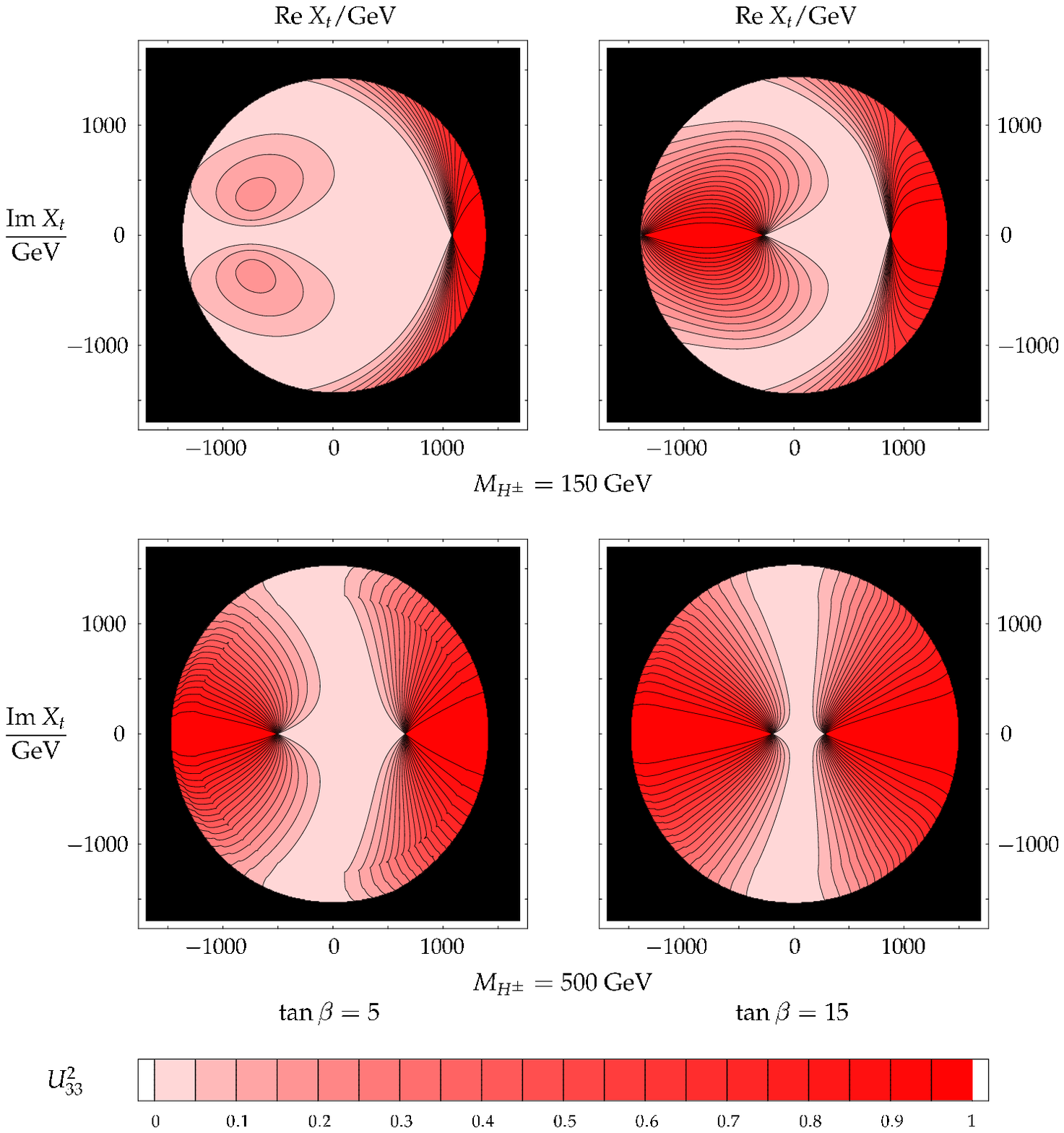}}
\caption{$U_{33}^2$ is shown in the $\re \XAt$--$\im \XAt$ plane for
$\tb = 5$ (left) and $\tb = 15$ (right) for $\mHp = 150 \gev$ (upper
row) and $\mHp = 500 \gev$ (lower row).
The other parameters are as given in \refeq{parameters}.
}
\label{fig:32mix}
\vspace{1em}
\end{figure} 

Now we focus on the mixing of the heavy neutral Higgs bosons in the presence
of complex parameters. 
As a measure of the mixing between these two states we show in
\reffi{fig:32mix} $U_{33}^2$ in the $\re\XAt$--$\im\XAt$~plane.
The choice for $\mHp$ and $\tb$ is the same as in \reffi{fig:DeM32At}, and the
other parameters are specified in \refeq{parameters}.
We have checked that $U_{13}^2$ is very close to zero (i.e.\ the lightest Higgs
boson is nearly a pure $\cp$-even state) and 
$U_{33}^2 \approx 1 - U_{23}^2$. The mixing varies strongly with
$\varphi_{\XAt}$ for both values of $\tb$ and low and high $\mHp$. 
In particular, for relatively large values of $|\XAt|$, 
the variation of the phase $\varphi_{\XAt}$ 
(with $|\XAt|$ kept fixed) can cause $U_{33}^2$ (and consequently also
$U_{23}^2$) to take on any value in the range $0 \leq U_{33}^2 \leq 1$.
It can furthermore be seen in all four panels of \reffi{fig:32mix}
that a large mixing between the two heavy neutral Higgs bosons, 
corresponding to the parameter regions where $U_{33}^2 \sim 0.5$, 
is a feature that can happen quite easily in the MSSM with complex
parameters. Studying the properties of the heavy Higgs bosons is therefore
of particular interest, since they could, at least in principle, give
access to large $\cp$-violating effects.

The connection between $\De M_{32}$ (the mass difference between the two heavy
Higgs bosons) and $U_{33}^2$ (the mixing of $\Hz$ and $\Hd$, i.e.\ the 
$\cp$~composition of the two heavy Higgs bosons) can be 
analyzed by comparing \reffi{fig:32mix} with \reffi{fig:DeM32At}
(the choice for $\mHp$ and $\tb$ is the same in both figures).
The regions of the nodal points in \reffi{fig:32mix}, i.e.\ the points in 
which a change in $\XAt$ causes the largest variation in $U_{33}^2$,
coincide with the regions where the mass difference $\De M_{32}$ is
close to zero. This behaviour, which occurs for all $\mHp$ and $\tb$
values, is clearly a resonance-type effect: in the parameter regions 
where the masses of the Higgs states become degenerate, the mixing
effects between the states are maximal.

\begin{figure}[htb!]
\centerline{\includegraphics{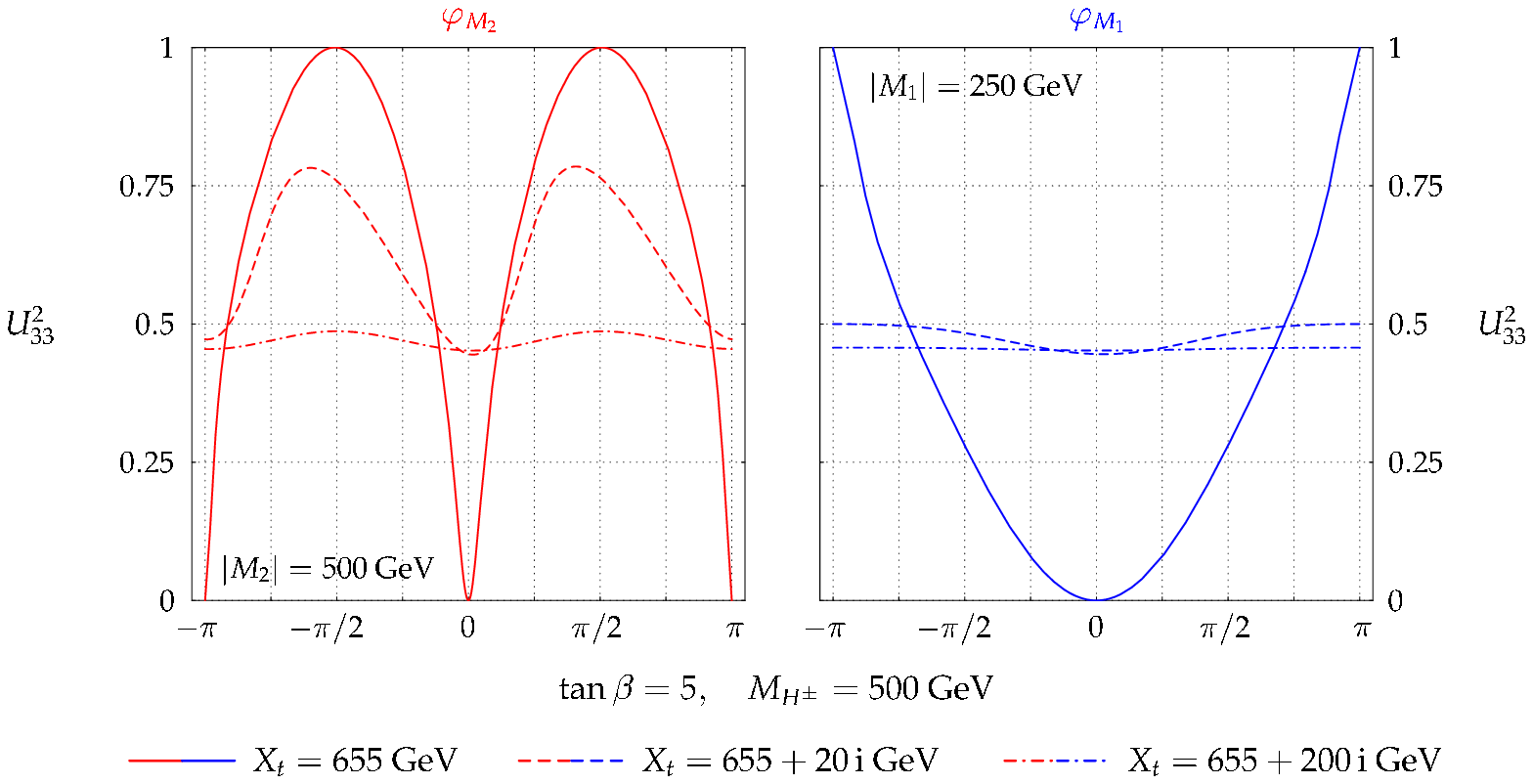}}
\caption{$U_{33}^2$ is shown as a function of $\phiMz$ (left plot) and $\phiMe$
  (right plot) for $\re \XAt = 655 \gev$ and 
$\im\XAt = 0, 20, 200 \gev$ (solid, dashed, dot-dashed lines). The other
  parameters are the same as in \reffi{fig:32mix}.
}
\label{fig:32mixphi12}
\end{figure}

Finally we analyze the effect of the gaugino phases on the mixing of the
heavy neutral Higgs bosons. 
The dependence of $U_{33}^2$ on the gaugino phases is depicted in
\reffi{fig:32mixphi12} for $\tb = 5$ and $\mHp = 500 \gev$.
We have chosen $\re\XAt$ as $\re\XAt = 655 \gev$ such that 
$\im\XAt = 0$ corresponds to a resonance region where $\De M_{32}$ is
close to zero, namely the right nodal point in the lower left plot of
\reffi{fig:32mix}. For displaying the effects of the gaugino phases
$\phiMz$ and $\phiMe$ we have chosen in \reffi{fig:32mixphi12} three
different values of the imaginary part of $\XAt$, 
$\im\XAt = 0, 20, 200 \gev$ (solid, dashed, dot-dashed lines).
For the case of the nodal point where $\im\XAt = 0$ a very strong variation 
of $U_{33}^2$ with both $\phiMz$ (left plot) and $\phiMe$ (right plot) 
is observed, covering the whole allowed range $0 \leq U_{33}^2 \leq 1$.
It should be noted that one encounters in this case 
a strong variation of $U_{33}^2$ with the gaugino phases even if
$\phiMz$, $\phiMe$ are restricted to relatively small values.
However, for $\im\XAt = 20 \gev$, i.e.\ only
slightly away from the nodal point, the dependence of $U_{33}^2$ on the
gaugino phases is already much smaller. For
$\im\XAt = 200 \gev$ the variation of $U_{33}^2$ with $\phiMz$ and $\phiMe$ is
numerically insignificant.
It becomes apparent that the gaugino
phases can have a strong impact on the mixing of the heavy Higgs
bosons, but only directly on resonance. Outside of the resonance regions
the effects of the gaugino phases are small.
This is in contrast to \citere{mhiggsCPXsn}, where it was claimed that 
a strong dependence of the Higgs mixing on $\phiMz$ and $\phiMe$ were
a general feature of the cMSSM.

\clearpage
\newpage
\section{The code \fhtt: program features}
\label{sec:feynhiggs}

\fhtt~\cite{mhiggslong,feynhiggs,feynhiggs1.2,feynhiggs2} is a Fortran
code for the evaluation of the masses, decays and production processes 
of Higgs bosons in the MSSM with real or complex
parameters. In this section we give a short overview about its
features. More detailed information about installation and use can be
found in the Appendix.

The calculation of the higher-order corrections is based on the
Feynman-diagrammatic (FD) approach as outlined in the previous sections.
At the one-loop level, it consists of a complete evaluation, including the
full momentum and phase dependence, and as a further option the full 
$6 \times 6$ non-minimal flavor violation (NMFV) 
contributions for scalar quarks~\cite{mhiggsNMFV,bsgNMFV}. At the
two-loop level all available corrections from the real MSSM have been
included (see~\citeres{mhiggsAEC,habilSH,mhiggsAWB} for reviews). They are
supplemented by the resummation of the leading effects from the
(scalar)~$b$ sector including the full complex phase
dependence~\cite{complexDeltab}. 

The loop-corrected pole masses are determined as the real parts of the
complex poles as described in \refse{subsec:massmix}. The imaginary
parts of the Higgs-boson self-energies are fully taken into account.
The masses are
evaluated with two independent numerical algorithms. Deviations
between the two methods indicate potential problems of numerical
stability. 
In addition to the Higgs-boson masses, the program also provides
results for the effective Higgs-boson couplings
and the wave function normalization factors 
for external Higgs bosons as described in
\refses{subsec:extHiggs},\ref{subsec:effcoupl}. 

Besides the computation of the Higgs-boson masses, effective couplings and 
wave function normalization factors,
the program also evaluates an estimate for 
the theory uncertainties of these quantities due to unknown
higher-order corrections. The total uncertainty is the sum of
deviations from the central value, $\Delta X = \sum_{i = 1}^3
|X_i - X|$ with $X = \{M_{h_1,h_2,h_3,H^\pm}, U_{ij}, Z_{ij}\}$, where
$U_{ij}$ is defined in \refeq{eq:defU} and $Z_{ij}$ in
\refeqs{eq:defZ}--(\ref{eq:zfactors123}). Alternatively instead of
$U_{ij}$ also $R_{ij}$, defined in \refeq{eq:defR}, can be evaluated.
The $X_i$ are given by
\begin{itemize}
\item
$X_1$ is obtained by varying the renormalization scale (entering via
the $\overline{\text{DR}}$ renormalization) within
$\frac 12 m_t \leqslant \mu \leqslant 2 m_t$,

\item
$X_2$ is obtained by using $m_t^{\text{pole}}$ instead of the running 
$m_t$ in the two-loop corrections,

\item
$X_3$ is obtained by using an unresummed bottom Yukawa coupling,
$y_b$, \ie a $y_b$ including  
the leading \order{\als\alb} corrections, but not resummed to 
all orders.
\end{itemize}

Furthermore \fhtt\ contains the evaluation of all relevant Higgs-boson
decay widths and effective couplings (the latter are given
in the conventions used in the MSSM model file of the program
\fa~\cite{feynarts}). In particular, the following quantities are
calculated:
\begin{itemize}
\item
the total width for the neutral and charged Higgs bosons,

\item
the branching ratios and effective couplings of the three neutral Higgs
bosons to 
\begin{itemize}
\item SM fermions (see also \citeres{hff,habilSH}),
      $h_i \to \bar f f$,
\item SM gauge bosons (possibly off-shell),
      $h_i \to \gamma\gamma, ZZ^*, WW^*, gg$,
\item gauge and Higgs bosons,
      $h_i \to Z h_j$,
      $h_i \to h_j h_k$,
\item scalar fermions,
      $h_i \to \tilde f^\dagger \tilde f$,
\item gauginos,
      $h_i \to \tilde\chi^\pm_k \tilde\chi^\mp_j$,
      $h_i \to \tilde\chi^0_l \tilde\chi^0_m$,
\end{itemize}

\item 
the branching ratios and effective couplings 
of the charged Higgs boson to 
\begin{itemize}
\item SM fermions,
      $H^- \to \bar f f'$,
\item a gauge and Higgs boson,
      $H^- \to h_i W^-$,
\item scalar fermions,
      $H^- \to \tilde f^\dagger \tilde f'$,
\item gauginos,
      $H^- \to \tilde\chi^-_k \tilde\chi^0_l$.
\end{itemize}

\item
the production cross sections 
of the neutral Higgs bosons at the Tevatron and the LHC in the
approximation where the corresponding SM cross section is rescaled 
by the ratios of the corresponding partial widths in the MSSM and the SM
or by the wave function normalization factors for external Higgs
bosons as defined through 
\refeqs{eq:defZ}--(\ref{eq:zfactors123}), see \citere{HiggsXS}
for further details.

\end{itemize}

\noindent
For comparisons with the SM, the following quantities are also evaluated
for SM Higgs bosons with the same mass as the three neutral MSSM Higgs
bosons:
\begin{itemize}
\item
the total decay width,

\item
the couplings and branching ratios of a SM Higgs boson to SM fermions,

\item
the couplings and branching ratios of a SM Higgs boson to 
SM gauge bosons (possibly off-shell).

\item
the production cross sections at the Tevatron and the
LHC~\cite{HiggsXS}.
\end{itemize}

\noindent
\fhtt\ furthermore provides results for electroweak
precision observables that give rise to constraints on the 
SUSY parameter space (see Ref.~\cite{PomssmRep} and references
therein):
\begin{itemize}

\item
the quantity $\De\rho$ up to the two-loop level~\cite{deltarho} that
can be used to
indicate disfavored scalar top and bottom mass combinations, 

\item
an evaluation of $\MW$ and $\sweff$, where the SUSY contributions are 
treated in the $\De\rho$ approximation (see e.g.\ \citere{PomssmRep}),
taking into account at the one-loop level the effects of complex phases 
in the scalar top/bottom sector~\cite{MWweber} as well as NMFV
effects~\cite{mhiggsNMFV},

\item
the anomalous magnetic moment of the muon, including a full one-loop
calculation~\cite{g-2MSSMf1l,g-2early} as well as leading and
subleading two-loop corrections~\cite{g-2FSf,g-2CNH} (see also
\citere{g-2Doink}),  

\item
the evaluation of $\br(b \to s \ga)$ including NMFV effects~\cite{bsgNMFV}. 
\end{itemize}

\noindent
Finally, \fhtt\ possesses some further features:
\begin{itemize}
\item
Transformation of the input parameters from the \drbar\ to the
on-shell scheme (for the scalar top and bottom parameters), including 
the full \order{\alpha_s} and \order{\alpha_{t,b}} corrections.

\item
Processing of SUSY Les Houches Accord (SLHA 2)
data~\cite{slha,slhaTH,slha2TH}.  
\fhtt\ reads the output of a spectrum generator file and evaluates the
Higgs boson masses, branching ratios etc.  The results are written in
the SLHA format to a new output file.

\item
Predefined input files for the SPS~benchmark scenarios~\cite{sps} and 
the Les Houches benchmarks for Higgs boson searches at hadron 
colliders~\cite{benchmark} are included.

\item
Detailed information about all the features of \fhtt\ are provided in
man pages.
\end{itemize}


\section{Conclusions}
\label{sec:conclusions}

We have presented 
new results for the complete one-loop contributions to the masses and 
mixing effects in the Higgs-boson sector of the MSSM with
complex parameters. They have been obtained in the
Feynman-diagrammatic approach using a hybrid renormalization scheme
where the masses are renormalized on-shell, while the \drbar\ scheme
is applied for $\tb$ and the field renormalizations. A detailed
description has been given of the renormalization procedure and of the 
determination of the masses as the real parts of the complex poles 
of the higher-order corrected Higgs propagator matrix.
Besides the Higgs-boson masses we have also derived the wave
function normalization factors needed for processes with external Higgs
bosons. We have discussed different ways for defining effective
Higgs-boson couplings that incorporate leading higher-order effects.
As a result, we propose effective couplings based on the ``$p^2$~on-shell'' 
approximation, where the
Higgs-boson self-energies are evaluated at the tree-level masses.

In our calculation of the Higgs-boson masses, couplings and wave
function normalizations the full dependence on all relevant complex phases is 
taken into account.
We incorporate for the first time the complete effects
arising from the imaginary parts of the one-loop Higgs-boson 
self-energies in a consistent way. Our result for the complete one-loop
contributions in the
cMSSM is supplemented by all available two-loop corrections in the
rMSSM and a resummation of the leading effects from the
sbottom sector for complex parameters.

In our numerical discussion we have first analyzed the impact of our
results on the physics of the light Higgs boson, which is of interest
in view of the current exclusion bounds and possible high-precision
measurements of the properties of a light Higgs boson at the next
generation of colliders. We first investigated the impact of the
different MSSM sectors and the possible effects of the corresponding
complex phases on the mass of the lightest neutral Higgs boson of
the cMSSM Higgs, $\MHe$, and the coupling of the lightest Higgs to gauge
bosons. 
The well-known result from the rMSSM
that the bulk of the corrections to $\MHe$ arises from the
fermion/sfermion sector is of course also reflected in the dependence on the
associated complex phases. We find that the
effects associated with the variation of $\varphi_{\Xt}$ are in general 
numerically very important, leading to shifts in $\MHe$ of up to 
$8 \gev$ in the examples that we have studied.
The corrections beyond the fermion/sfermion loops, arising from the 
chargino/neutralino sector, the gauge-boson
sector and the Higgs sector, can amount up to about $3 \gev$. The
dependence of $\MHe$ on the gaugino phases $\phiMz$ and $\phiMe$ is in
general rather small and will be difficult to resolve even in the
high-precision environment of the ILC (in particular in view of the
existing experimental constraints on the gaugino phases).

We have furthermore compared our full result with various aproximations. 
We find sizable deviations of up to $1.5 \gev$ in $\MHe$ between the
full result and the ``$p^2 = 0$'' approximation, which is often used in the
literature. We find that the ``$p^2$~on-shell'' approximation is
significantly closer to the full result with maximum
deviations below $0.5 \gev$.

For the example of the partial decay widths of the three neutral Higgs 
bosons into $\tau$ leptons we have compared the results based on the wave
function normalization factors of the external Higgs bosons with
effective-coupling approximations. The dependence of the partial widths 
on the phase $\varphi_{\Xt}$ is very pronounced. We find that the result
based on an effective coupling in the ``$p^2 = 0$'' approximation
(corresponding to the effective potential approach) deviates from the
full result by typically up to 5\%, with maximum deviations of more than
10\%. The effective couplings in the ``$p^2$~on-shell'' approximation 
show a better agreement with the full result for not too large values
of $\MHp$.

While over a large part of the cMSSM parameter space the lightest
neutral Higgs boson is almost a pure $\cp$-even state, large
$\cp$-violating effects may influence the masses and mixings of the two
heavier neutral Higgs bosons of the cMSSM. We have analyzed the impact
of complex phases on the mass difference between the two heavy Higgs
bosons, $\De M_{32}$, and on their mixing properties. Our full result has 
been compared with various approximations. We find that the mass
difference $\De M_{32}$ can significantly be enhanced by threshold
effects, so that mass differences of more than $10 \gev$ are possible
even in the decoupling region where $\mHp \gg \MZ$. Since the threshold
effects go beyond the ``$p^2 = 0$'' approximation, large deviations
between the full result and this approximation may occur. The
``$p^2$~on-shell'' approximation, on the other hand, is close to the
full result even in the threshold region. We have furthermore shown that
effects of the imaginary parts of the Higgs-boson self-energies can be
sizable in the threshold region. They can be as large as about $5 \gev$
in this region.

While the mass difference $\De M_{32}$ is sensitive to the effects of 
the complex phases in the sfermion sector (the effects of the gaugino
phases, on the other hand, are small), a determination of $\De M_{32}$
will in general not be sufficient to establish the existence of non-zero
complex phases. We have shown that most values of $\De M_{32}$ that can
be obtained in the complex $\re \Xt$--$\im \Xt$ plane can also be
realized on the real axis, i.e.\ for $\im \Xt = 0$. In order to extract
information on the complex phases, the mass difference $\De M_{32}$ will
have to be combined with a suitable set of observables that exhibit a
non-trivial dependence on the complex phases. 

We find that a large ($\cp$-violating) mixing between the two heavy Higgs 
states is possible over a significant part of the cMSSM parameter space.
In the parameter regions where the mass difference $\De M_{32}$ becomes
very small a resonance-type behaviour is possible. It gives rise to large
variations in the mixing between the two Higgs bosons, i.e.\ a
small change in the phase $\varphi_{\Xt}$ can have a dramatic effect on
the mixing properties. Directly on resonance even the gaugino phases 
$\phiMz$ and $\phiMe$ have a large impact on the Higgs mixing.
In contrast to  \citere{mhiggsCPXsn}, where it was claimed that
a strong dependence of the Higgs mixing on $\phiMz$ and $\phiMe$ were
a general feature of the cMSSM, we find that outside of the resonance
regions the effects of the gaugino phases are small. For a reliable
description of the resonance region it is crucial to correctly take into
account the imaginary parts of the Higgs-boson self-energies. 

Our analysis has shown that effective couplings in the 
``$p^2 = 0$'' approximation, as used in the effective potential
approach, can be insufficient for correctly matching the
Higgs-mixing properties to the higher-order corrected Higgs-boson
masses. The effective couplings
based on the ``$p^2$~on-shell'' 
approximation that we have studied in this paper, on the other hand,
have turned out to be well-suited for a numerical description of the
Higgs-boson mixing for not too large values of $\MHp$. However, 
for large $\MHp$ the correct
decoupling properties of the effective couplings of the lightest Higgs
boson 
are achieved in the ``$p^2 = 0$'' approximation, but not in the
``$p^2$~on-shell'' approximation.

Our results for the Higgs-boson masses, couplings and wave function
normalization factors together with an estimate of the remaining
theoretical uncertainties from unknown higher-order corrections
are implemented into the public Fortran
code \fhtt. The code also contains the evaluation of the Higgs-boson
decays and the main Higgs-boson production channels at the Tevatron and 
the LHC, calculated using the full wave function normalization
factors. 
Further quantities that are useful for deriving constraints on the SUSY
parameter space are also evaluated, such as
electroweak precision observables, the anomalous magnetic moment of the muon 
and (in the case of complex parameters) electric dipole moments.
The code can be obtained from {\tt www.feynhiggs.de}\,.


\subsection*{Acknowledgements}
We thank K.~Williams for numerous checks and helpful discussions.
We are grateful to D.~St\"ockinger for illuminating discussions, in
particular on \citere{susyren}, and we
also thank S.~Hesselbach, C.~Schappacher and P.~Slavich for
interesting discussions. 
W.H.\ wants to express special thanks to the Institute of
Theoretical Physics, University of Vienna, where this
paper was finalized while he was Erwin Schr\"odinger visiting
professor.


\newpage
\begin{appendix}

\section{The code \fhtt: installation and use}

\subsection{Installation and Use}

The installation process is straightforward and should take no more than 
a few minutes:
\begin{itemize}
\item Download the latest version from \texttt{www.feynhiggs.de} and 
      unpack the tar archive.

\item The package is built with \texttt{./configure} and \texttt{make}.
      This creates the library \texttt{libFH.a} and the command-line 
      frontend \texttt{FeynHiggs}.

\item To build also the Mathematica frontend \texttt{MFeynHiggs},
      invoke \texttt{make all}.

\item \texttt{make install} installs the files into a platform-dependent
      directory tree, for example \texttt{i586-linux/\{bin,lib,include\}}.

\item Finally, remove the intermediate files with \texttt{make clean}.
\end{itemize}
\fhtt\ has four modes of operation,
\begin{itemize}
\item Library Mode:
Invoke the \fh\ routines from a Fortran or C/C++ program linked against
the \texttt{libFH.a} library.

\item Command-line Mode:
Process parameter files in native \fh\ or SLHA format at the shell
prompt or in scripts with the standalone executable \texttt{FeynHiggs}.

\item WWW Mode:
Interactively choose the parameters at the \fh\ User Control Center 
(FHUCC) and obtain the results on-line.

\item Mathematica Mode:
Access the \fh\ routines in Mathematica via MathLink with 
\texttt{MFeynHiggs}.
\end{itemize}


\subsubsection{Library Mode}

The core functionality of \fhtt\ is implemented in a static Fortran 77 
library \texttt{libFH.a}.  All other interfaces are `just' frontends to 
this library.

In view of Fortran's lack of symbol scoping, all internal symbols have
been prefixed to make symbol collisions very unlikely.  Also, the
library contains only subroutines, no functions, which simplifies the
invocation.  In Fortran, no include files are needed except for access
to the coupling structure.  In C/C++, a single include file
\texttt{CFeynHiggs.h} must be included once for the prototypes.
Detailed debugging output can be turned on at run time.

The library provides the following functions:
\begin{itemize}
\item
\texttt{FHSetFlags} sets the flags for the calculation.

\item
\texttt{FHSetPara} sets the input parameters directly, or \\
\texttt{FHSetSLHA} sets the input parameters from SLHA data.

\item 
\texttt{FHSetCKM} sets the elements of the CKM matrix.

\item
\texttt{FHSetNMFV} sets the off-diagonal soft SUSY-breaking parameters
in the scalar quark sector that induce NMFV effects.

\item
\texttt{FHSetDebug} sets the debugging level.

\item
\texttt{FHGetPara} retrieves (some of) the MSSM parameters calculated
from the input parameters, \eg the sfermion masses.

\item
\texttt{FHHiggsCorr} computes the corrected Higgs masses, effective
couplings and wave function normalization factors.

\item
\texttt{FHUncertainties} estimates the uncertainties of the Higgs 
masses, effective couplings and wave function normalization factors.

\item
\texttt{FHCouplings} computes the Higgs couplings and BRs.

\item
\texttt{FHHiggsProd} calculates the Higgs-boson production
cross-sections at the Tevatron and the LHC.

\item
\texttt{FHConstraints} evaluates further electroweak precision 
observables.
\end{itemize}
These functions are described in detail in their respective man pages
in the \fh\ package. 


\subsubsection{Command-line Mode}

The \texttt{FeynHiggs} executable is a command-line frontend to the
\texttt{libFH.a} library.  It is invoked at the shell prompt as
\begin{verbatim}
   FeynHiggs inputfile [flags [scalefactor]]
\end{verbatim}
where
\begin{itemize}
\item \texttt{inputfile} is the name of a parameter file (see below).
\item \texttt{flags} is an (optional) string of integers giving the
  flag values, 
      \eg \texttt{40030211} (for details see the description of 
      \texttt{FHSetFlags} in the man pages). If
      \texttt{flags} is not specified, \texttt{40020211} is used.
The fifth flag controls the evaluation of the effective couplings,
  where \texttt{0} correspongs to the "$p^2$~on-shell" and \texttt{4}
  to the "$p^2 = 0$" approximation.
\item \texttt{scalefactor} is an optional factor multiplying the
      renormalization scale.  It is used to determine the dependence on 
      the renormalization scale, \eg by varying \texttt{scalefactor}
      from \texttt{0.5} to \texttt{2}.
\end{itemize}
\texttt{FeynHiggs} understands two kinds of parameter files:
\begin{itemize}
\item
Files in SUSY Les Houches Accord (SLHA) format.  In this case \fh\ adds 
the Higgs masses, mixings and decay widths to the SLHA data structure
and writes the latter to a file \textit{inputfile}\texttt{.fh}.

In fact, \fh\ tries to read each file in SLHA format first, and if that
fails, falls back to its native format.

\item
Files in its native format, for example
\begin{verbatim}
  MT         171.4
  MB         4.7
  MW         80.392
  MZ         91.1875
  MSusy      500
  MA0        200
  Abs(M_2)   200
  Abs(MUE)   1000
  TB         5
  Abs(Xt)    1000
  Abs(M_3)   800
\end{verbatim}
The syntax should be pretty self-explanatory.  Complex quantities can be 
given either in terms of absolute value \texttt{Abs(X)} and phase 
\texttt{Arg(X)}, or as real part \texttt{Re(X)} and imaginary part 
\texttt{Im(X)}.
Abbreviations, summarizing several parameters (such as \texttt{MSusy})
can be used, or detailed information about the various soft
SUSY-breaking parameters can be given.

Furthermore, it is possible to define loops over parameters, to scan 
parts of parameter space.  For example,
\begin{verbatim}
  TB         5 25 5
  MA0        100 800 *2
  MSusy      500 1000 /3
\end{verbatim}
declares three loops: \\
a) over $\tb$ from 5 to 25 linearly in steps of 5
   (\ie 5, 10, 15, 20, 25), \\
b) over $\MA$ from 100 to 800 logarithmically in steps of 2
   (\ie 100, 200, 400, 800), \\
c) over $\msusy$ from 500 to 1000 linearly in 3 steps
   (\ie 500, 750, 1000).

The output is written in a human-readable form to the screen. Since this
can be quite lengthy, a \% is printed in front of all lines with
`non-essential' information, \eg the details on couplings and decay
widths.  Thus to display only the `essential' information, one just has 
to ``grep'' away the \% lines, \ie
\begin{verbatim}
  FeynHiggs inputfile flags | grep -v %
\end{verbatim}
The output can also be piped through the \texttt{table} filter to yield 
a machine-readable version appropriate for plotting etc.  For example,
\begin{verbatim}
  FeynHiggs inputfile flags | table TB Mh0 > outputfile
\end{verbatim}
creates \texttt{outputfile} with two columns, $\tan\beta$ and $\Mh$.
The syntax of the output file is given as screen output.
\end{itemize}
Debugging output is governed by the environment variable 
\texttt{FHDEBUG} which can be set to an integer from 0 to 3 (for 
details see the description of \texttt{FHSetDebug} in the man pages).
For example,
\begin{verbatim}
   setenv FHDEBUG 1     (in csh or tcsh)
   export FHDEBUG=1     (in sh or bash)
\end{verbatim}
sets debugging level 1.


\subsubsection{WWW Mode}

The \fh\ User Control Center (FHUCC) is a WWW interface to the 
command-line executable \texttt{FeynHiggs}.  It provides a convenient 
way to play with parameters, but is of course not suited for large-scale 
parameter scans or extensive analyses.

To use the FHUCC, point your favorite Web browser at
\begin{verbatim}
   http://www.feynhiggs.de/fhucc
\end{verbatim}
adjust the parameters, and submit the form to see the results.  At the 
end of the result page, the input file used for that \fh\ run is 
presented, too.


\subsubsection{Mathematica Mode}

The \texttt{MFeynHiggs} executable provides access to the \fh\ functions
from Mathematica via the MathLink protocol.  This is particularly
convenient both because \fh\ can be used interactively this way and
because Mathematica's sophisticated numerical and graphical tools, \eg
FindMinimum, are available.

After starting Mathematica, install the package with
\begin{verbatim}
   In[1]:= Install["MFeynHiggs"]

   Out[1]= LinkObject[./i586-linux/bin/MFeynHiggs, 1, 1]
\end{verbatim}
which makes all \fh\ subroutines available as Mathematica functions.  
For details of their use, see the corresponding man pages.


\end{appendix}


\end{document}